\documentclass[a4paper, 12pt]{article}  

\usepackage{amssymb}
\usepackage{amsmath}
\usepackage[english]{babel}
\usepackage[nottoc]{tocbibind}
\usepackage{graphicx}
\usepackage{mathptmx}
\usepackage{makecell}
\usepackage{listings}
\usepackage{xcolor}
\usepackage{geometry}

\definecolor{codegreen}{rgb}{0,0.6,0}
\definecolor{codegray}{rgb}{0.5,0.5,0.5}
\definecolor{codepurple}{rgb}{0.58,0,0.82}
\definecolor{backcolour}{rgb}{0.95,0.95,0.92}

\lstdefinestyle{mystyle}{
    backgroundcolor=\color{backcolour},   
    commentstyle=\color{codegreen},
    keywordstyle=\color{magenta},
    numberstyle=\tiny\color{codegray},
    stringstyle=\color{codepurple},
    basicstyle=\ttfamily\footnotesize,
    breakatwhitespace=false,         
    breaklines=true,                 
    captionpos=b,                    
    keepspaces=true,                 
    numbers=left,                    
    numbersep=5pt,                  
    showspaces=false,                
    showstringspaces=false,
    showtabs=false,                  
    tabsize=2
}

\graphicspath{{./Images/}}

\newtheorem{theorem}{Theorem}[section]

\newtheorem{definition}{Definition}[section]

\def\logo{{\includegraphics[width=32mm]{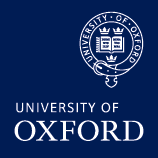}}}

\DeclareMathOperator{\img}{img}
\DeclareMathOperator{\id}{id}

\newcommand{\submittedtext}{{A dissertation submitted for the degree of}}
\newcommand{\dissertationtitle}{{From augmented microscopy to the topological transformer: a new approach in cell image analysis for Alzheimer's research}}
\renewcommand{\maketitle}{%
    \renewcommand{\footnotesize}{\small}
    \renewcommand{\footnoterule}{\relax}
    \thispagestyle{empty}
  \null\vfill
  \begin{center}
    { \Huge {\bfseries {\dissertationtitle}} \par}
{\large \vspace*{20mm} {\logo \par} \vspace*{10mm}}
    {{\Large Wooseok Jung} \par}
\vspace*{1ex}
    {\Large University of Oxford \par}
\vspace*{10mm}
    {{\submittedtext} \par}
\vspace*{1ex}
    {\it {Master of Mathematics in Mathematics} \par}
\vspace*{2ex}
    {{Trinity 2021}} \\
  \end{center}
  \null\vfill
}

\newenvironment{romanpages}
{\cleardoublepage\setcounter{page}{1}}
{\cleardoublepage\setcounter{page}{1}}

\begin{document}

\baselineskip=18pt plus1pt

\setcounter{secnumdepth}{7}
\setcounter{tocdepth}{4}

\maketitle                  
\newpage
\begin{abstract}
\thispagestyle{empty}  
Cell image analysis is crucial in Alzheimer's research to detect the presence of A$\beta$ protein inhibiting cell function. Deep learning speeds up the process by making only low-level data sufficient for the fruitful inspection. We first found Unet is most suitable in augmented microscopy by comparing performance in multi-class semantics segmentation. We develop the augmented microscopy method to capture nuclei in a brightfield image and the transformer using Unet model to convert an input image into a sequence of topological information. The performance regarding Intersection-over-Union is consistent concerning the choice of image preprocessing and ground-truth generation. Training model with data of a specific cell type demonstrates transfer learning applies to some extent.

The topological transformer aims to extract persistence silhouettes or landscape signatures containing geometric information of a given image of cells. This feature extraction facilitates studying an image as a collection of one-dimensional data, substantially reducing computational costs. Using the transformer, we attempt grouping cell images by their cell type relying solely on topological features. Performances of the transformers followed by SVM, XGBoost, LGBM, and simple convolutional neural network classifiers are inferior to the conventional image classification. However, since this research initiates a new perspective in biomedical research by combining deep learning and topology for image analysis, we speculate follow-up investigation will reinforce our genuine regime. 
\end{abstract}
\newpage

\begin{romanpages}          
\tableofcontents            
\end{romanpages}    


\section{Introduction}

Alzheimer's disease (AD) is known as the most common cause of dementia, one of the disorders that has not conquered yet. Nearly 850,000 people in the UK suffer from AD, and the number will increase to one million in 2025 \cite{prince2014dementia}. Amyloid-$\beta$-peptide (A$\beta$) composes amyloid plagues abundantly found in Alzheimer's disease patients \cite{Esch1122}, and is widely understood to contribute to the development of the disease. For instance, this toxic protein is known for impairing neuronal activities by reducing the number of activated synapses and causing calcium ion channels malfunction \cite{Arispe1993}. Hence, diminishing the level of $A\beta$ is the core of current AD treatments, although none of them can completely cure the disease \cite{Mark6239}. 

Phenotypic screening of target cells aims to test the efficacy of a candidate protein inhibiting $A\beta$ \cite{swinney2013contribution}. If a chemical alleviates $A\beta$-driven damage, there will be visible perturbations in cell-level features. Cell Painting \cite{bray2016cell} is a morphological profiling method accessible to organelles using different fluorescent dyes. Unlike conventional screening methods, cell painting can capture multiple cell components, more than three or four, at once by staining channel by channel. Therefore, it is suitable for large-scale experiments. 

CellProfiler  \cite{carpenter2006cellprofiler} is a fluorescent microscopy image analysis software that quantifies morphological features, including shape, size, location, and count. It uses classical machine learning algorithms to analyze data from cell painting \cite{carpenter2006cellprofiler}.  It facilitates thresholding or edge detection rather than modern techniques, including deep learning-based image classification and segmentation. Therefore, although CellProfiler performs feature extraction tasks with high fidelity, it is not suitable for further analysis for latent features. Beyond the old-school machine learning techniques, biomedical image analysis demands a new paradigm: deep learning. 
\begin{figure}[tbhp]
    \centering
    \includegraphics[width = \linewidth]{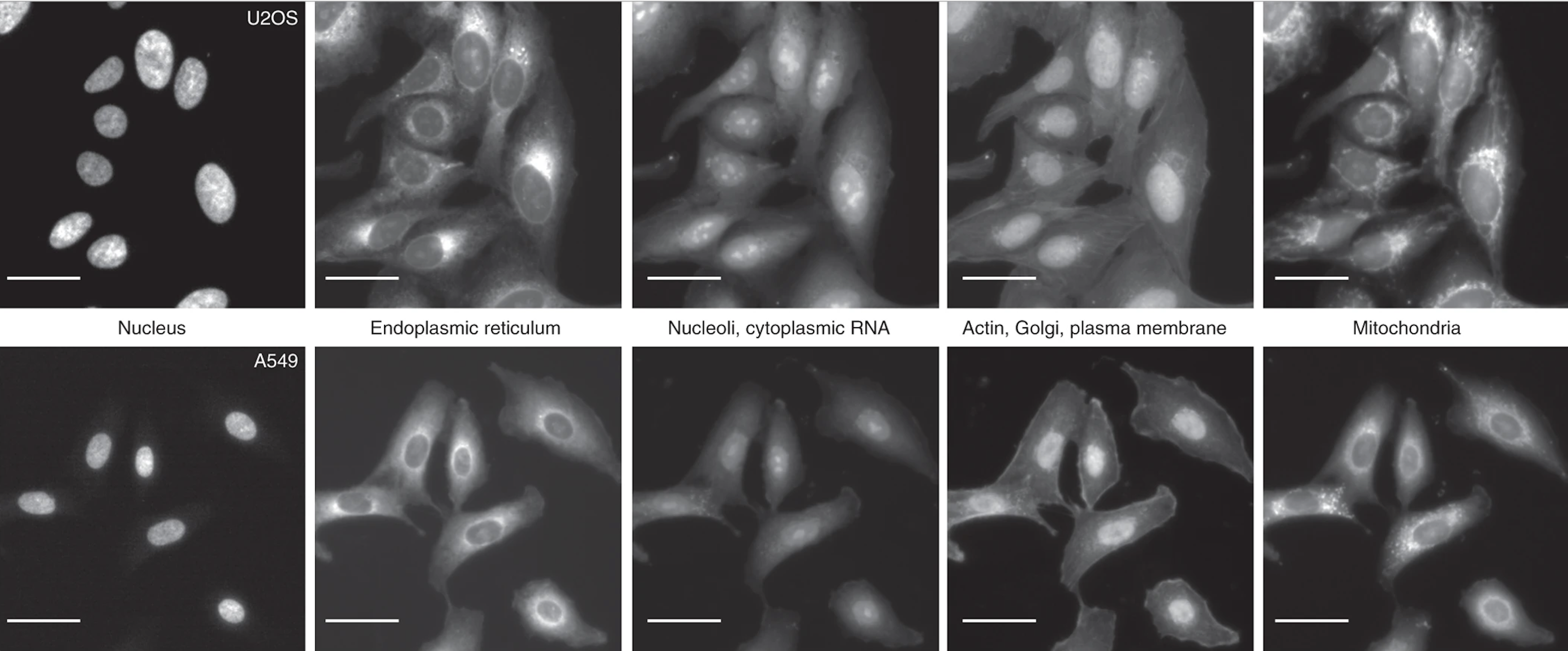}
    \caption{The Cell Painting assay in two different types of cells: U2OS and A549. Each fluorescent dyes reveals corresponding cell components, colored in bright white in each column. Figure from \cite{bray2016cell}.}
    \label{figcellpainting}
\end{figure}

There are thousands of potential candidates suppressing the effect of $A\beta$, but it is costly and inefficient to examine them one by one. High-throughput screening (HTS) allows scientists to carry out tests with speed and accuracy where deep learning takes the integral role. Deep neural networks guarantee effective and efficient research, including discovering unknown compounds with antibacterial activity \cite{STOKES2020688}, predicting compound activity through image analysis of assays \cite{SIMM2018611,MoenErick2019Dlfc}, and undergoing cell type classification and cell-level morphological phenotyping \cite{Yao2019}. 

Topological data analysis (TDA) is a state-of-the-art framework to access the shape of a complex dataset. Topological inference in statistics takes advantage of its deterministic environment, while deep learning is often unstable and prone to overfitting \cite{chazal2021introduction}. TDA is not only applied for feature engineering \cite{kovacev2016using,li2014persistence} but also used for model selection \cite{carlsson2020topological}. TDA fortifies biomedical research by analyzing morphological transition of cell configuration reacting from a specific treatment, for instance. 

This dissertation aims to design novel machine learning methods for cellular image analysis in Alzheimer's disease research. We start by reviewing fundamental concepts in deep learning and basic semantic segmentation models. Then, we survey topological data analysis with concepts in algebraic topology and its extension to computational fields. In section 4, we first share results in deep-learning-based multiclass image segmentation with different network models. Also, we examine the augmented microscopy method to segment out nuclei from a brightfield image. We evaluate the transferability of both approaches. In topological data analysis simulation, we implement two topological transformers, the silhouette and the signature transformers, which can extract topological features from the 2D grayscale image and convert them into one-dimensional sequences. Applying these transformers, we compare fluorescent nuclei image classification performances of various machine learning models to a traditional convolutional neural network. In the end, we discuss how our deep learning and topological data analysis methods bring an impact on biomedical research and a potential pipeline to synthesize the augmented microscopy and TDA-based image classification. All important codes are provided in Appendix A.

\subsection{Related Works}
CellProfiler is prominent in studying the size or intensity of cell images \cite{bray2015using}. Since it relies on a general machine learning scheme, CellProfiler is not limited to a few types of cells. It can evaluate treatment performance of leukaemias and lymphomas through cancer cell image analysis \cite{snijder2017image}, examine virus structure \cite{sakurai2015two}, and portrait growth of bacteria \cite{stanley2014identification}. There are some variations of the CellProfiler. CellProfiler 3.0 \cite{mcquin2018cellprofiler} can work on 3D images, CellProfiler Analyst \cite{jones2008cellprofiler} for high-dimensional images containing complex features , and CellProfiler Tracer \cite{bray2015cellprofiler} for a time-dependent dataset. 

Deep learning facilitates biomedical researches based on substantial amounts of data. One main application of it is cellular image analysis \cite{MoenErick2019Dlfc}. A neural network can classify images of cells showing morphological transition \cite{10.1093/bioinformatics/btx069} or partition the images into substructures and label each structure \cite{10.1007/978-3-319-24574-4_28,chen2016combining}. 
DeepCell \cite{van2016deep} is a biological image segmentation library which enables biology laboratories to access deep learning more conveniently. Caicedo et al (2019) \cite{caicedo2019evaluation} shows deep learning outperforms CellProfiler in nucleus segmentation tasks, especially Unet and a convolutional neural network based on DeepCell. Also, instead of a monochrome brightfield image, one can set up the augmented microscopy by stacking fluorescent channels and performing \textit{in silico} labelling \cite{ChristiansenEricM2018ISLP}. Recent research about applying deep learning for automated cell painting experiment to capture signatures from Parkinson's disease patients shows the deep learning model can point out that phenotypes differ individually \cite{schiff2020deep}.

Topological data analysis in medical research is relatively new but still used in different fields. Topological descriptors are used in classification of breast cancer type \cite{singh2014topological}, traumatic brain injury \cite{nielson2015topological}, fMRI images \cite{stolz2020topological}. Not only image analysis but TDA is also applied in studying molecular structures \cite{pike2020topological} or relationship to protein functionality \cite{dorier2018knoto}. 
In most researches, TDA makes unsupervised pattern detection to reveal novel signatures in data, meaning that TDA is used for feature extraction, followed by applying for statistical machine learning. We formulate classification of topological features through neural networks like \cite{umeda2017time} but for images instead. 

\subsubsection{Softwares}
Gudhi \cite{maria2014gudhi} is a famous Python-based libraries for topological data analysis, and TDAstats \cite{wadhwa2018tdastats} is an R-based package for computing persistent homology. Although computing persistent homology is a computationally expensive task, the development of the discrete Morse theory relaxes it. Gudhi, the one we use mainly, can construct a complex, draw a persistence diagram, produce a persistence landscape, and compute metrics between barcodes. Those barcodes are our primary interest, which is the input of classification tasks. See section 3.3 for more details about the fundamentals of topological data analysis and how these software work.

\section{Deep Learning and Image Segmentation}
Deep learning is a 'black box model' composed of many layers computing multiple matrix operations. Deep learning methods consist of two steps before training: constructing a sequence of layers and selecting a proper optimization method. 

\subsection{Fundamentals of Deep Learning}

\subsubsection{Types of Layers}
By increasing a number of layers and units within a layer, a neural network can take into account more complex task \cite{Goodfellow-et-al-2016}. Let $h^{0} = \textbf{x}$ be a representation of the input $\textbf{x}$. Then following representations can be recursively defined by composition of layers $f^{(i)}$ which is 
\begin{equation}
    \begin{split}
        h^{l} &= f^{(l)}(h^{l-1}), \\
        f(\textbf{x}) &= h^{m} = f^{(m)} \circ f^{(m-1)} \circ \cdots f^{(1)}(\textbf{x})
    \end{split}
\end{equation}
where $h^{m}$ is the output layer.   
A most elementary type of layer is a fully-connected layer or simply linear module. It takes the input vector $\textbf{x} \in \mathbb{R}^{n}$ and print output $f(x) \in \mathbb{R}^{m}$ by
\begin{equation}
    f(\textbf{x}) = W\textbf{x} + \textbf{b}
\end{equation}
where $W \in \mathbb{R}^{m \times n}$ and $\textbf{b} \in \mathbb{R}^{m}$ are weights and biases respectively. If there are multiple linear modules in a network, then each modules takes an input that is the output of the preceding linear module.

However, using only linear modules is sometimes insufficient to demonstrate the complexity of the model. Thus, we may add some nonlinear units after a linear layer. These units are called activation layers and have an integral role in the computation, increasing the performance of a deep feedforward network model. Followings are the three most popular choices:
\begin{enumerate}
    \item rectified linear unit: $\text{ReLU}(x) = \max{\{0, x\}}$ 
    \item tanh: $\tanh{(x)} = \exp{(x)} - \exp{(-x)} / \exp{(x)} + \exp{(-x)}$
    \item sigmoid: $\text{sigmoid}(x) = 1/(1 + \exp{(-x)})$.
\end{enumerate}
\begin{figure}[tbhp]
    \centering
    \includegraphics[width = \linewidth]{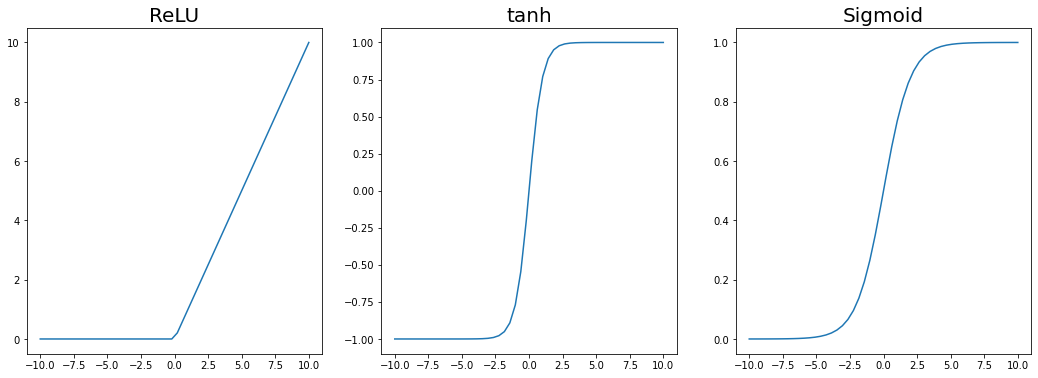}
    \caption{Plots of ReLU, tanh, and sigmoid activation units.}
    \label{figact}
\end{figure}
Figure \ref{figact} shows a plot of the three hidden units. These are component-wise functions, so it returns the output vector with the same dimension as the input vector. Most modern deep learning models prefer the ReLU unit since it preserves non-negative values and passes more information to the next layer unlike the other two \cite{Goodfellow-et-al-2016}. Also, the tanh unit is known to perform better in general than sigmoid \cite{Goodfellow-et-al-2016}. First, it is closer to the identity function: tanh(0) = 0 while sigmoid(0) = 0.5. This implies when tanh takes the value near zero, it behaves like a linear function, making a network to be trained easier. 

If a model requires to print a probability vector, such as image classification task whose output vector corresponds to probability to be assigned to each class, softmax layer is added at the end:
\begin{equation}
    \text{softmax}([x_{1}, \dots x_{n}]) = 
    \left[ \frac{\exp{(x_{1})}}{\sum_{i = 1}^{n}\exp{(x_{i})}}, \dots ,  \frac{\exp{(x_{n})}}{\sum_{i = 1}^{n}\exp{(x_{i})}}\right].
\end{equation}
Note that the softmax layer is not element-wise. Multi-layer perceptron (MLP) is a concatenation of multiple linear modules followed by activation layers, which is the most fundamental structure of a deep feedforward network. 

Although MLP is simple and intuitive, it does not fit in every deep learning task. Since it collapses the structure of the input, it is not suitable for the dataset like images where the location information of each entry, or pixel, is significant \cite{VALUEVA2020232}. Therefore, starting from the classic LeNet \cite{lecun1989backpropagation,lecun1999object}, convolutional neural network (CNN) is widely used in many tasks, mainly in computer vision. Unlike the linear module, a convolutional layer has a fixed size of the kernel, sliding on the input matrix (or vector if its one-dimensional CNN). Each convolutional layer has a kernel with fixed parameters, therefore, instead of optimizing $m \times n$ parameters in a linear module, a convolutional layer requires much fewer parameters. 
\begin{equation}
    \begin{split}
         \text{module} &\sim \text{Conv2d}(c_{in}, c_{out}, d_{1}, d_{2}) \\
         x^{out} & = \text{module}(x^{in}) \\
         x^{out}_{i',j',k'} & = \sum_{i=1}^{d_{1}}\sum_{j = 1}^{d_{2}}\sum_{k = 1}^{c_{in}}w_{i,j,k,k'} \cdot x_{i'+i-1, j'+j-1, k}^{in} \hspace{0.5cm} \forall i', j', k'.
    \end{split}
\end{equation}
The $d_{1} \times d_{2}$ matrix $(w_{ij})$ is a kernel, and every input-output channel pair $(k, k')$ has its own kernel. 

\begin{figure}[tbhp]
    \centering
    \includegraphics[width = \linewidth]{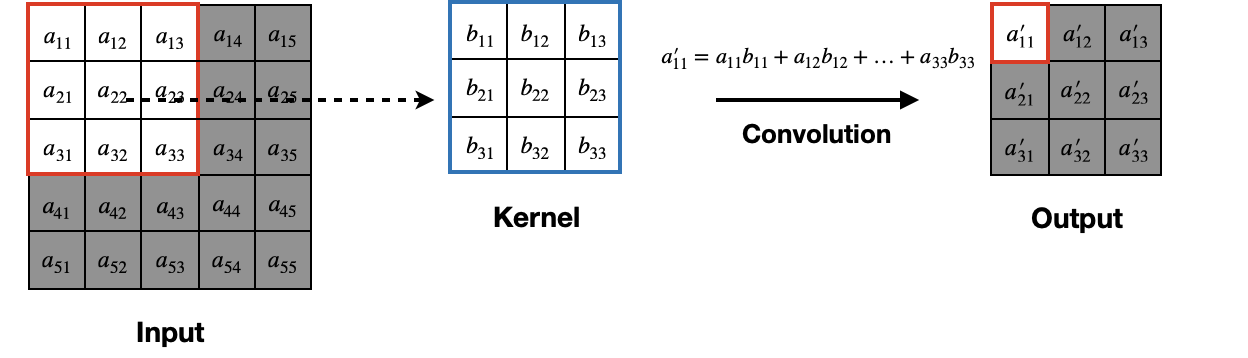}
    \caption{An illustration of a convolutional layer mapping $5 \times 5$ input to $3 \times 3$ output with $3 \times 3$ kernel, no padding and stride 1. The kernel is placed on the input, compute convolution at its position, and locate the result value in the output. It slides on the whole input to compute remaining convolution values.}
    \label{figconv}
\end{figure}

Some additional variations are available. In Figure \ref{figconv}, the kernel starts with the input center at $a_{22}$. We can add zero padding, surrounding the input with 0's to make the convolution start at $a_{11}$, for instance. $P-$padding determines how thick the zeros. If the input in Figure $\ref{figconv}$ get 1-padding, then the resultant input will be $7 \times 7$. Stride determines the step of sliding the kernel. 1-stride makes the kernel moves to adjacent center pixel $(a_{22}$ to $a_{23})$, but if the stride is 2, then the kernel jumps to center at $a_{24}$. Indeed, an input or kernel need not to be a square. Hence, in the 2D convolution, an input image with size $(H_{in}, W_{in})$ with Kernel size $(d_{1}, d_{2})$, padding $(P_{1}, P_{2})$, stride $(S_{1}, S_{2})$ transforms to an output with size $(H_{out}, W_{out})$ such that 
    \begin{align*}
         H_{out} &= \frac{H_{in} + 2P_{1} - d_{1}}{S_{1}}+1 \\
        W_{out} &= \frac{W_{in} + 2P_{2} - d_{2}}{S_{2}}+1.
    \end{align*}

Convolutional layers reduce the size of the image and increase the number of channel in general. On the contrary, deconvolutional, or convolutional transpose, layers do the exact opposite thing. Up-sampling through deconvolutional layers is necessary in generative tasks, for instance DC-GAN \cite{radford2015unsupervised} or semantic segmentation \cite{long2015fully}.

\begin{figure}
    \centering
    \includegraphics[width = 0.8\linewidth]{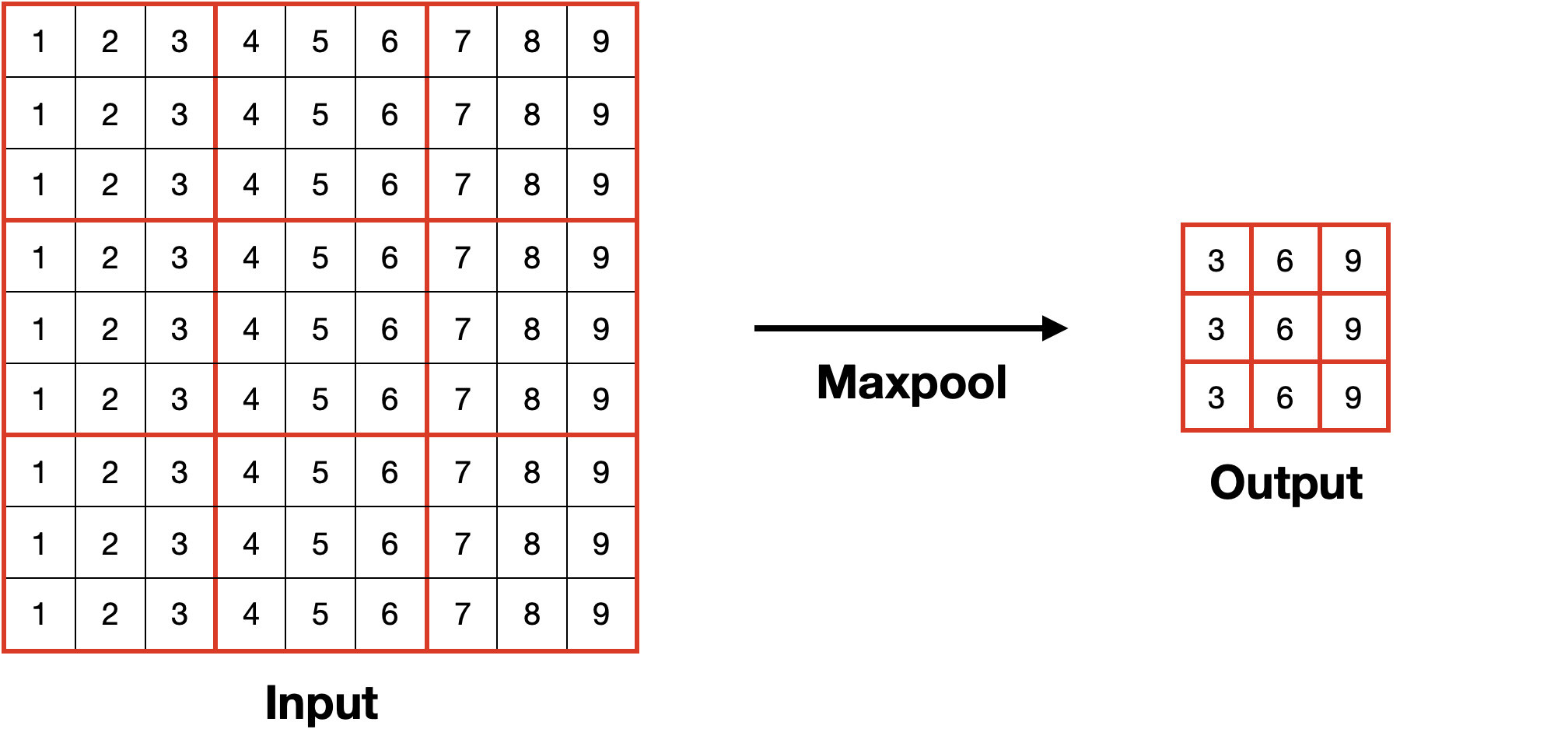}
    \caption{Example of a maxpooling layer, mapping $9 \times 9$ input to $3 \times 3$ output.}
    \label{figMaxpool}
\end{figure}

Instead of strided convolutions, one can use pooling layer. For example, maxpooling layer \cite{zhou1988image} takes the maximum value of over elements of a region with fixed size:
\begin{equation}
    \begin{split}
       x^{out} &= \text{Maxpool}_{d_{1}, d_{2}}(x^{in}) \\
    x^{out}_{i',j',k'} &= \max_{i = 1}^{d_{1}}\max_{j = 1}^{d_{2}}x_{i+d_{1}(i'-1),j+d_{2}(j'-1),k'}.  
    \end{split}
\end{equation}
Figure \ref{figMaxpool} shows how maxpooling works with given input and pooling size. Unlike the convolutional layer, a pooling layer partitions the input and not channel-dependent. Minimum or average value is also applicable as an alternative of maximum.   

Batch normalization layers \cite{pmlr-v37-ioffe15} reparametrize layers to reduce complexity \cite{Goodfellow-et-al-2016}. While training, the distribution of each layer keep changes, requiring it to adapt a new distribution every time \cite{pmlr-v37-ioffe15}. Batch normalization solves this internal covariance shift through normalizing each dimension with respect to mean and variance from each mini-batch. So, let $\textbf{B} = \{B_{1}, \dots B_{m}\}$ be a mini-batch of size $m$. The transformation replaces $B$ with 
\begin{equation*}
    \textbf{B}' = \frac{\textbf{B} - \mu}{\sigma}
\end{equation*}
where $\mu = \frac{1}{n}\sum_{i}B_{i}$ and $\sigma = \sqrt{\epsilon + \frac{1}{n}\sum_{i}(B_{i} - \mu)^{2}}$. The $\epsilon$ is a small constant to prevent the denominator becoming zero. At the end, instead of the raw normalized output $\textbf{B}'$, batch normalization exploits affinely transformed version $\gamma \textbf{B} + \beta$ with learnable parameters $\gamma$ and $\beta$ to keep the network demonstration. 

\subsubsection{Optimization}
Assume each $f^{(i)}$ in (1) is differentiable and have paramter vector $\theta^{(i)}$. So the inclusive function $f$ has parameter vector $\theta = [\theta^{1}, \dots \theta^{m}]$. Then, optimizing a neural network is equivalent to the following empirical risk minimization task.
\begin{equation}
    \min_{\theta}\hat{R}(\theta), \hspace{2cm} \hat{R}(\theta) = \lambda r(\theta) + \frac{1}{n}\sum_{i = 1}^{n}L(y_{i}, f_{\theta}(x_{i}))
\end{equation}
where $r$ is a regularizer and $\lambda$ is a predefined regularization strength. Starting from the initial assumption of $\theta$ $\theta_{0}$, gradient descent step forms 
\begin{equation}
    \theta_{t} = \theta_{t-1} - \alpha\nabla_{\theta}\hat{R}(\theta_{t-1}),
\end{equation}
where $\alpha$ is the step size and $n$ is the size of the training set. Then we get the recusrive form of $\frac{\partial L_{i}}{\partial \theta^{(l)}}$ which is 
\begin{equation}
    \frac{\partial L_{i}}{\partial \theta^{(l)}} =  \frac{\partial L_{i}}{\partial h_{i}^{m}} \frac{\partial h_{i}^{m}}{\partial h_{i}^{m-1}} \cdots \frac{\partial h_{i}^{l+1}}{\partial h_{i}^{l}} \frac{\partial h_{i}^{l}}{\partial \theta^{(l)}}
\end{equation}
where each $\displaystyle\frac{\partial h_{i}^{k}}{\partial h_{i}^{k-1}}$ represents the Jacobian matrix of $f^{(k)}$ differentiated by $h_{i}^{(k-1)}$. 

The backward computation, starting from $\frac{\partial L_{i}}{\partial h_{i}^{m}}$, is less costly than the computation with reversed direction. This is called \textit{backpropagation}. Although backpropagation is computationally lighter, it requires more memory.

$n$ in (6) represents the size of training set. However, computing the global gradients every step is impractical. Instead, Stochastic Gradient Descent (SGD) method facilitates randomly-chosen minibatches are used in the computation of the gradient. So, let $B = {x_{1}, \dots x_{m}}$ be a randomly chosed subset of the training set with $|B| \ll n$. Then the stochastic estimate of the gradient is 
\begin{equation}
    \widehat{\nabla_{\theta}\hat{R}(\theta)} = \lambda \frac{\partial r}{\partial \theta} + \frac{1}{|B|}\sum_{x \in B}\frac{\partial L_{i}}{\partial \theta} \vert_{y, f(x)}.
\end{equation}
So the dataset is split into $n / m$ mini-batches. After running each epoch, the mini-batch selection is initialized and computed again. Note that the model is not always globally convex: there is a possibility to reach local minima, not global. 

The choice of the step size, or learning rate, $\alpha$ is significant in optimization. Too large $\alpha$ makes the optimization unstable, but too small $\alpha$ leads to very slow convergence. Instead of manually defining the step size, learning schedulers make the step size varies in the training process. It determines the current state and makes an appropriate update: reduce the step size to prevent SGD from being too variant, or increase if the process is far away from the optimum.

For example, the cosine annealing scheduler makes the learning rate decreases with a fixed rule. Let $\eta_{max}$ be the initial learning rate, $\eta_{min}$ be the predifined lower bound, and $T_{cur}$ be the number of epochs passed. Then $\eta_{t}$ is given as follows:
\begin{equation}
    \eta_{t} = \eta_{min} + \frac{1}{2}(\eta_{max} - \eta_{min})(1 + \cos{\frac{T_{cur}}{T_{max}}\pi}).
\end{equation}
\subsection{Image Segmentations Models}

Image segmentation partitions a given image into multiple parts. Unlike image classification which assigns an image to its label, image segmentation classifies each pixel, so it can define boundaries or locate an object. Image segmentation is widely used in self-driving cars \cite{treml2016speeding}, tumor detection in lung X-ray \cite{nam2020undetected}, and satellite images \cite{bhandari2014cuckoo}.

There are mainly two types of image segmentation. Semantic segmentation is the way to detect the 'class' of each image. So even if there are multiple images with the same class, they are classified as the same. On the other hand, instance segmentation identifies each pixel to the instance of an object, so even if two different objects are in the same class, instance segmentation regards them as distinct. In our research, we only do semantic segmentation tasks.

Now we introduce some most famous models for semantic segmentation. 

\subsubsection{Fully Convolutional Networks}
As its name, a fully convolutional network consists of convolutional layers only, followed by a deconvolutional layer as the output layer. Since fully connected layers collapse information related to the location of each pixel, replacing them with convolutional layers is more reliable in the context of image segmentation where positional information is crucial. 

\begin{figure}[tbhp]
    \centering
    \includegraphics[width = \linewidth]{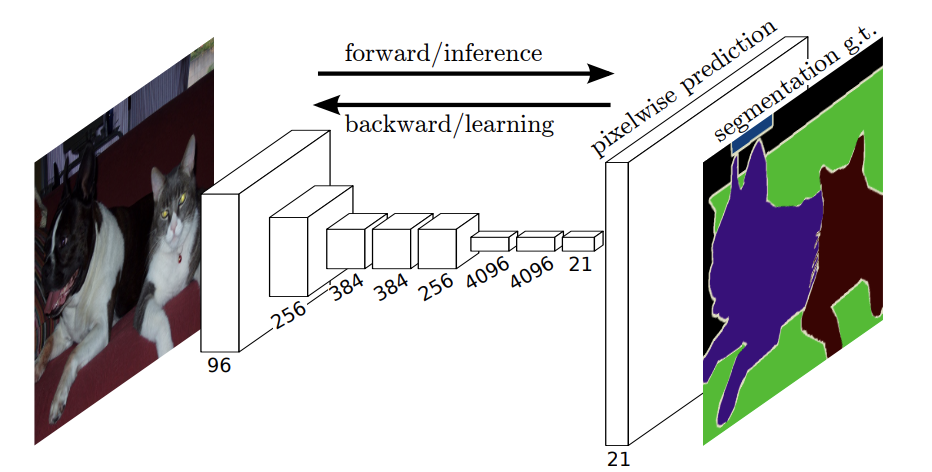}
    \caption{Visualization of a fully convolutional network. Figure from \cite{long2015fully}}
    \label{figFCN}
\end{figure}

Instead of bilinear interpolation, FCN utilizes deconvolutional, or convolution transpose layers composed of learnable parameters. Figure \ref{figFCN} shows the pipeline of an FCN. After multiple convolutional layers, the data passes a single deconvolution layer. The deconvolution resolution can be modified, and finer deconvolution leads to more accurate segmentation.  However, performing a single up-sampling loses much information. Therefore, FCN uses skip connection to preserve features from shallow layers by adding them to deeper layers. Also, lowering the stride of deconvolution increases segmentation performance \cite{long2015fully}.

\subsubsection{The U-Net}

\begin{figure}[tbhp]
    \centering
    \includegraphics[width = \linewidth]{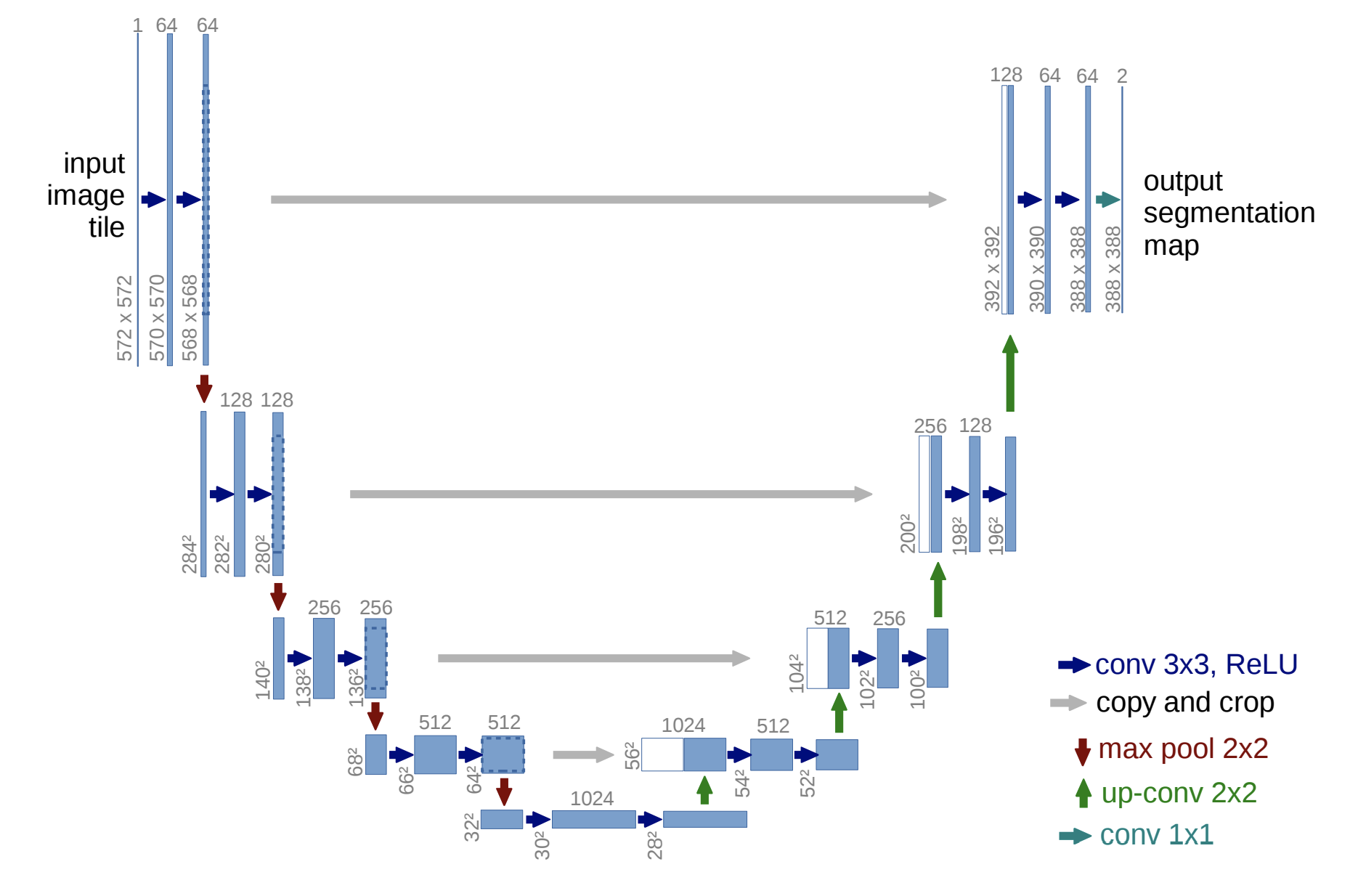}
    \caption{Visualization of the structure of the U-net. Figure from \cite{10.1007/978-3-319-24574-4_28}.}
    \label{figUnet}
\end{figure}

U-net \cite{10.1007/978-3-319-24574-4_28} is initially developed for biomedical image analysis. It admits a fully convolutional structure but has multiple up-sampling blocks and concatenates the skip connections, unlike the FCN. The U-net consists of two main paths. In the left half of Figure \ref{figUnet}, downsampling blocks contains two $3 \times 3$ convolution layers followed by ReLU activation and max-pooling, doubling the number of channels. Also, results after convolution in each block are saved, concatenated to the corresponding upsampling block. 

On the other hand, in the right half, max-pool layers are replaced by deconvolution layers in the upsampling part, which factor the number of channels by two. Multiple upsampling blocks make a more detailed recovery of the input image. In binary segmentation, each pixel of the ground truth is 0 or 1: 0 refers to the background, and 1 refers to the object. So, each pixel in the output of the U-net is equivalent to a 2-dimensional vector showing probability to be assigned to the corresponding channel. For example, if a pixel in output has a value (0.7, 0.3), it means the pixel corresponds to the background with a probability of 0.7 and the object with a probability of 0.3. It implies channel-wise softmax layer includes at the end of the U-net. In Figure \ref{figUnet}, an output image is smaller than the input because convolutions in upsampling blocks have 0-padding, but this condition can be relaxed to make the output have the same size as the input. 

\subsubsection{DeepLabv3}

\begin{figure}
    \centering
    \includegraphics[width = \linewidth]{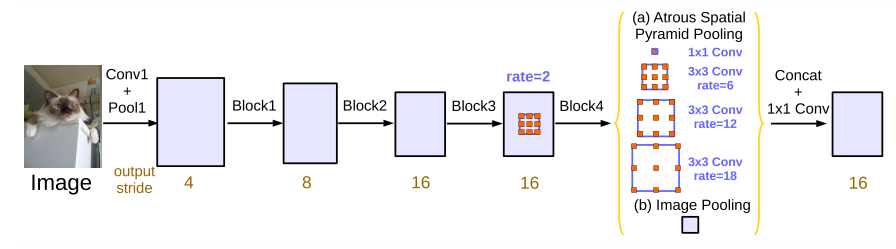}
    \caption{Module propagation in downsampling of DeepLabv3 with Atrous Spatial Pyramidal Pooling (ASPP). Figure from \cite{chen2017rethinking}.}
    \label{figdeeplabv3}
\end{figure}

Similar to the two previous models, DeepLabv3 \cite{chen2017rethinking} admits a convolutional upsampling structure. However, DeepLabv3 exploits atrous convolutions instead of deconvolutional layers in downsampling.
Given two-dimensional input $x$ and a learnable kernel $w$, each pixel $y[i]$ of the output $y$ is 
\begin{equation}
    y[i] = \sum_{k}x[i + r \cdot k]w[k]
\end{equation}
with stride rate $r$. $r = 1$ in the usual convolution, but when $r > 1$, it makes a sparse window of pixel as displayed in Figure \ref{figdeeplabv3}. Computing and stacking multiple atrous convolution layers with different rates, it returns Atrous Spatial Pyramida Pooling (ASPP) \cite{chen2017deeplab} which preserves more details than normal convolution followed by pooling \cite{chen2017rethinking}. Resnet, mainly Resnet-50 or Resnet-101 \cite{he2016deep}, is the default backbone of deeplabv3, whose spirit comes from the skip connection: adding features from previous convolutions to pertain low-level features to some level. 

\begin{figure}
    \centering
    \includegraphics[width = \linewidth]{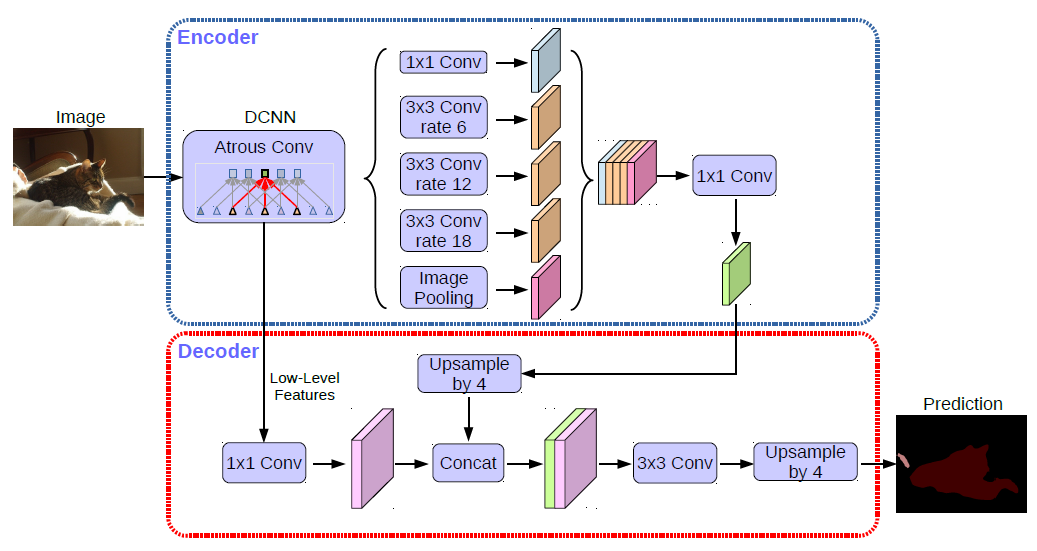}
    \caption{Diagram of DeepLabv3+ model which utilize DeepLabv3 in its encoder. Figure from \cite{chen2018encoderdecoder}}
    \label{figdeeplabv3+}
\end{figure}

Deeplabv3+ is an extension of DeepLabv3 plus encoder-decoder structure \cite{chen2018encoderdecoder}. DeepLabv3+ admits concatenation of low-level features to intermediate outputs in a decoder, as U-net did. So, module cascade in Figure \ref{figdeeplabv3} substitutes the encoder block in Figure \ref{figdeeplabv3+}. Figure \ref{figdeeplabv3+} shows the model applies $1 \times 1$ convolution to the low-level feature before the concatenation to compress the number of the channels. Compression to 48 or 32 channels is optimal, which is a trade-off between model complexity and preserving features \cite{chen2018encoderdecoder}. 

\section{Topological Data Analysis}
Topological Data Analysis (TDA), first introduced in \cite{carlsson2009topology}, is a geometrical method to access latent structure of the data \cite{chazal2021introduction}. Suppose we have two sets of data points sampled from $\mathbb{R}^{2}$. We need to know the shape of the underlying distribution to determine geometric difference between the datasets. Calculating pointwise mean squared error does not work if the numbers of data points are different, and a point pairing rule is well-defined in general. TDA solves the problem by detecting holes and cliques of the data using concepts in algebraic topology. For example, a disc and an annulus are different since the latter has a hole, but the former does not. 

TDA views the data as a point cloud embedded in an ambient metric space, such as $\mathbb{R}^{n}$ with the Euclidean metric. Then it requires a cascade of the data to demonstrate a continuous change of the underlying topology. TDA extracts quantified topological features from each step and uses them to describe the shape of the data.
\subsection{Fundamentals}
Constructing a cascade of the given data demands a sequence of simplicial complexes. Let $X = \{x_{0}, \dots x_{m}\}$ be a given dataset, a subset of a metric space $(M, d)$. Then a simplicial complex $K$ of $X$ is a subset of the power set $P(X)$ such that every singleton subset of $X$ is in $K$, and every subset of an element in $K$ belongs to $K$. Since $K$ is not uniquely determined unless $X$ is a singleton, there exist several methods to construct a simplicial complex from $X$.
\begin{definition}[Vietoris-Rips Complex]
The Vietoris-Rips complex $VR_{\epsilon}(X)$ is a set of simplices $\sigma \in P(X)$ such that $d(\alpha, \beta) < \epsilon$ for all $\alpha, \beta \in \sigma$. 
\end{definition}

\begin{definition}[\u{C}ech Complex]
The \u{C}ech Complex $C_{\epsilon}$ is a set of simplices $\sigma$ such that $\bigcap_{\alpha \in \sigma}\bar{B}_{\epsilon}(\alpha) \neq \emptyset$, where $\bar{B}_{\epsilon}(\alpha)$ denotes the closed ball centered at $\alpha$ with radius $\epsilon$.
\end{definition}

\begin{figure}
    \centering
    \includegraphics[width = \linewidth]{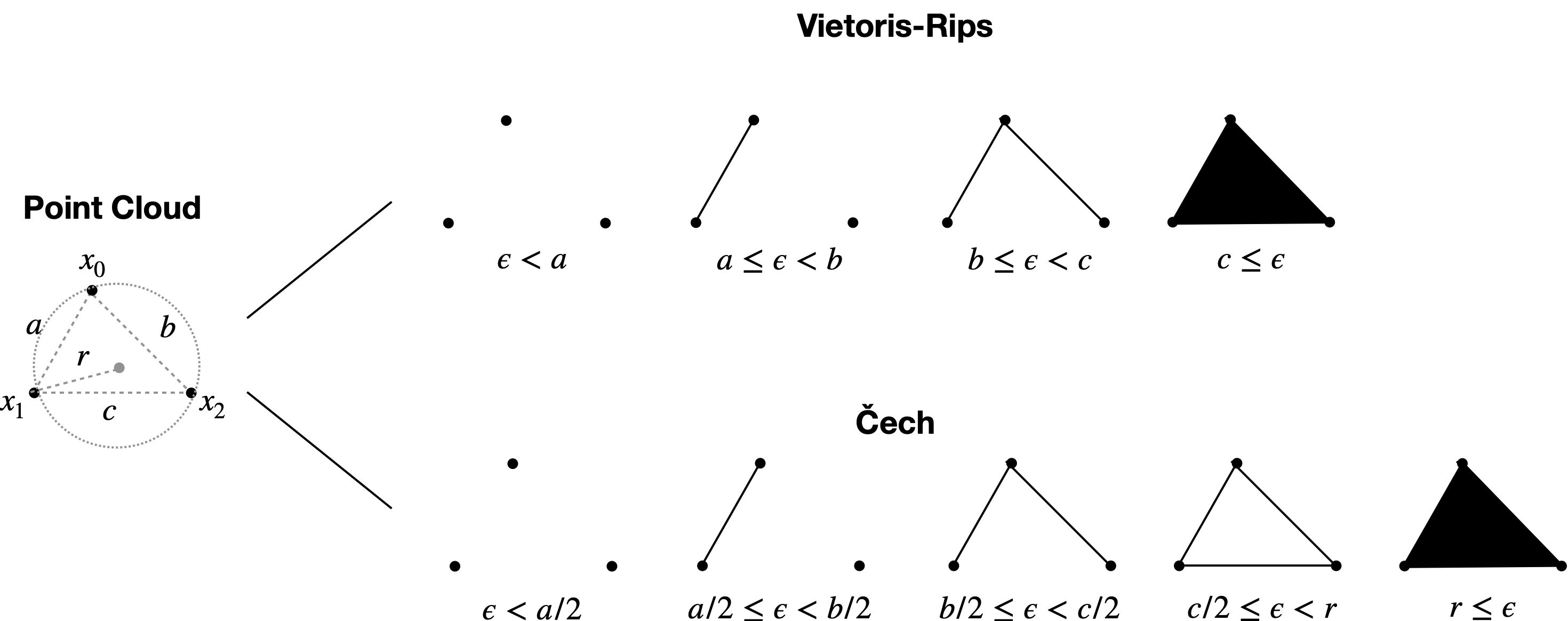}
    \caption{An example point cloud $X = \{x_{0}, x_{1}, x_{2}\}$ and its corresponding Vietoris-Rips and \u{C}ech complexes for each $\epsilon$. $a, b$, and $c$ denote lengths of 1-simplices and $r$ is the radius of the circumscribed circle.}
    \label{VRCech}
\end{figure}

Figure \ref{VRCech} displays a point cloud $X$ consists of three data points and its two different complex constructions. In Vietoris-Rips complex, after two 1-simplices arising, it immediately jumps to the filled triangle, the solid 2-simplex. If all $\{x_{0}, x_{1}\}$, $\{x_{1}, x_{2}\}$, and $\{x_{0}, x_{2}\}$ are in $VR_{\epsilon}(X)$, then $\{x_{0}, x_{1}, x_{2}\}$ must be in the complex either. However, in \u{C}ech complex, it passes the case of hollow 2-simplex. Also, the scale of $\epsilon$ is different: $\{x_{0}, x_{1}\}$ arises at $\epsilon = a$ in Vietoris-Rips but at $\epsilon = a/2$ in \u{C}ech. This gives the following inequality: 
\begin{equation}
    VR_{\epsilon}(X) \leq C_{\epsilon}(X) \leq VR_{2\epsilon}(X).
\end{equation}
Note that the closed balls of $C_{\epsilon}(X)$ forms a cover of $X$. The nerve theorem ensures that $C_{\epsilon}(X)$ is homotopic equivalent to the cover. 
\begin{theorem}[Nerve theorem]
Let ($U_{i}$) be a cover of $X$. Then the nerve $N(U_{\bullet})$ is the simplicial complex over the index set I of the cover where $\sigma \in U_{\bullet}$ if and only if 
$\textbf{Supp}(\sigma) \colon = \bigcap_{i \in \sigma}U_{i} \neq \emptyset$.
If $\textbf{Supp}(\sigma)$ is contractible for any simplex $\sigma$ in $N(U_{\bullet})$, then $|N(U_{\bullet})|$ is homotopy equivalent to X.
\end{theorem}
Given a cover consisting of closed balls centered at vertices with radius $\epsilon$, $C_{\epsilon}(X)$ is the nerve of the cover. It implies \u{C}ech complexes maintain the topological features of the data, which is not guaranteed for Vietoris-Rips Complex. On the other hand, Vietoris-Rips complexes only require pairwise distances, where \u{C}ech complexes need distances computed further, for instance $r$ in figure \ref{VRCech}. This enables Vietoris-Rips complexes compute topology much faster. 

The nested sequences $(VR_{i}(X))_{i \in \mathbb{R}_{+}}$ and $(C_{i}(X))_{i \in \mathbb{R}_{+}}$ induce filtrations of simplicial complexes. A filtration of a simplicial complex $K$ over $X$ is a nested sequence of subcomplexes of $K$ such that:
\begin{equation}
    X = K_{0} \leq K_{1} \leq K_{2} ... \leq K_{n} = K.
\end{equation}
We can naturally extend the indices to the non-negative reals. 

However, computing filtrations of both Vietoris-Rips and \u{C}ech complexes are inefficient for a large dataset. Both complexes access at most k-1-dimensional simplicial complexes if there are $k$ data points, even if the underlying topology is simple \cite{de2004topological}.
One modern alternative is the alpha complex \cite{edelsbrunner1993union}, a subcomplex of the Delaunay triangulation. 

\begin{definition}[Alpha Complex]
Let $V(x)$ be the Voronoi diagram of the point cloud $X$ containing $x \in X$. Let $R_{\epsilon}(x) = B_{\epsilon}(x) \cap V(x)$. Then the alpha complex $A_{\epsilon}(X)$ is defined by 
$A_{\epsilon}(X) = \{\sigma \in K \mid \bigcap_{x \in \sigma}R_{\epsilon}(x)\}$.
\end{definition}
Since $R_{\epsilon}(x) \subset B_{\epsilon}(x)$, the alpha complex is a subcomplex of the \u{C}ech complex. Since all partitions of the Voronoi diagram and closed balls are contractible, geometric realizations of the two complexes are homotopy equivalent. The alpha complex filtration is composed of simplicial complexes $(K_{i})$ where $K_{i} = A_{\sqrt{i}}(X)$.

Adapting the Delaunay triangulation exempts computing high-dimensional cliques, but it is efficient triangulation computation that is the problem \cite{de2004topological}.  An example to relieve the complexity is the witness complex \cite{de2004topological}, which diminishes the computational complexity by introducing landmark points, a relaxed version of the Delaunay triangulation. 

\subsubsection{Persistent Homology}
A filtration of simplicial complexes induces the sequence of homology groups. Persistent homology tracks the evolution of those homology groups. Homology is roughly a measure to detect cycles that are not a boundary of a manifold. For each $k \geq 0$, the $k$-th chain group of a simplicial complex $K$ is the vector space $C_{k}(K)$ over $\mathbb{F}$ spanned by the $k$-simplicies in $K$. Let $\sigma = \{v_{0}, \dots , v_{k}\}$ be a $k$-simplex in $K$. Assuming all simplices in K are ordered, the i-face of $\sigma$ is the $(k-1)-$simplex $\sigma_{-i}$ obtained by getting rid of $v_{i}$ from $\sigma$. For example, a 2-simplex $\{x_{0}, x_{1}, x_{2}\}$ in Figure \ref{VRCech} has three $i-$faces, each of them is precisely an edge of the triangle. This leads to the definition of the (algebraic) boundary.
\begin{definition}[Boundary]
Let $\sigma$ be a $k-$dimensional simplex in $K$. Then the boundary of $\sigma$ is 
\begin{equation}
    \partial_{k}(\sigma) = \sum_{i = 0}^{k}(-1)^{k}\sigma_{-i}.
\end{equation}
\end{definition}

Thus, the boudnary of $\sigma$ is in the image of the map $\partial_{k}^{K}:$ $C_{k}(K) \longrightarrow C_{k-1}(K)$. Therefore, we can construct a sequence of chain groups connected by boundary maps, such as
\begin{equation}
    \cdots C_{k}(K) \xrightarrow{\partial_{k}^{K}} C_{k-1}(K) \xrightarrow{\partial_{k-1}^{K}} \cdots \xrightarrow{\partial_{1}^{K}} C_{0}(K) \xrightarrow{} 0.
\end{equation}
The collection $(C_{*}(K), \partial_{*}^{K})$ is a chain complex if and only if $\partial_{k}^{K} \circ \partial_{k+1}^{K} = 0$ for all $k \geq 0$. \begin{definition}[Homology]
The $k-$th homology group of the chain complex \\
$(C_{*}(K), \partial_{*}^{K})$ is the quotient space
\begin{equation}
    H_{k}(C_{*}, \partial_{*}) = \ker{\partial_{k}} / \img{\partial_{k+1}}.
\end{equation}
\end{definition} 
The $k-$th homology groups of a given filtration defines a sequence of vector spaces with associated linear map $a_{i \rightarrow j}:$ $H_{k}(F_{i}K) \xrightarrow{} H_{k}(F_{j}K)$ induced from the simplicial inclusion map $\iota_{i \rightarrow j}: F_{i}K \hookrightarrow  F_{j}K$. The sequence $(V_{*}, a_{*}) = (H_{k}(F_{*}K), a_{*})$ is called a persistence module. Persistent homology group of a persistence module is given by
\begin{equation}
    PH_{i \rightarrow j}(V_{*}, a_{*}) = \img{a_{i \rightarrow j}}.
\end{equation}

Note that persistence homology familiarizes when a non-boundary loop arises and dies. For example, a $k$- dimensional loop is born at $t = t_{1}$ becomes a basis element of $H_{k}(F_{t}(K))$. If it disappears at $t = t_{2}$, then the map $a_{t_{1} \rightarrow t_{2}}$ maps the loop to zero. But for all $t_{3} < t_{2}$, $a_{t_{1} \rightarrow t_{3}}$ has a non-trivial image.

\subsubsection{Barcodes and Persistence Diagram}
The barcode of a persistence module carries information on the lifespans of the loops. For each pair of $0 \leq i \leq j$, the interval module $(I_{*}^{i,j}, c_{*}^{i,j})$ over a field $\mathbb{F}$ is given by
\begin{equation*}
  I_{k}^{i,j} =   
  \begin{cases}
  \mathbb{F} \hspace{0.3cm} \text{if} \hspace{0.1cm} i \leq k \leq j \\
  0 \hspace{0.3cm} \text{otherwise}
  \end{cases}
  \text{and}
  \hspace{0.25cm}
  c_{k}^{i,j} = 
  \begin{cases}
  \id_{\mathbb{F}} \hspace{0.3cm} \text{if} \hspace{0.1cm} i \leq k \leq j \\
  0 \hspace{0.3cm} \text {otherwise}
  \end{cases}.
\end{equation*}
So, an interval module is a bar of length $j - i$. The structure theorem allows the unique decomposition of any persistence module. 
\begin{theorem}[The structure theorem of a persistence module] \ \\
For any persistence module $(V_{*}, a_{*})$, there exists a set of intervals $B(V_{*}, a_{*})$ such that 
\begin{equation}
    (V_{*}, a_{*}) \cong \bigoplus_{[i,j] \in B(V_{*}, a_{*})}(I_{*}^{i,j}, c_{*}^{i,j})^{\mu(i,j)},
\end{equation}
where $\mu(i,j)$ is the multiplicity of the interval $[i,j]$. 
\end{theorem}

\begin{figure}[tbhp]
    \centering
    \includegraphics[width = 0.8\linewidth]{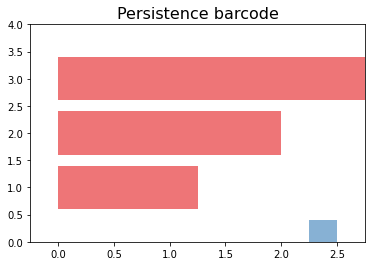}
    \caption{The barcode of a \u{C}ech filtration of the point cloud in Figure \ref{VRCech}. $x_{0}, x_{1}, x_{2}$ are embedded to points (0,0), (1,2), and (3,0) respectively. The list of points are converted to Simplextree data using Gudhi. The red and blue bars denote 0-dimemnsional and 1-dimensional persistence respectively.}
    \label{figbarcode}
\end{figure}

Figure \ref{figbarcode} shows the barcode of the \u{C}ech filtration in Figure \ref{VRCech}. Remark that the 0-dimensional homology group is equivalent to space spanned by one's connected components. At $t = 0$, there exists three connected components. So, $H_{0}(F_{0}K) = \mathbb{F}^{3}$. As $d(x_{0},x_{1}) = \sqrt{5}$, $\{x_{0},x_{1}\}$ is born at $ t = 1.25 = (\sqrt{5}/2)^{2}.$ $x_{0}$ component is 'absorbed' to the $x_{1}$, hence the bottom red bar ends. At $t = 2.25$, all the three 1-simplices are alive but not the 2-simplex. Hence, there exists a loop consisting of the edges of a triangle, and $ H_{1}$ is non-trivial. At $t = 2.5$, the interior of the triangle is filled, so the loop disappears. From the homology groups, the persistence homology $PH_{0 \rightarrow 1}$, for instance, is $\mathbb{F}^{2}$ since it maps $\mathbb{F}^{3}$ to $\mathbb{F}^{2}$ for which $x_{0}$ and $x_{1}$ are mapped to same basis element and $x_{2}$ to the other.

Measuring similarity between two barcodes requires a notion of metric. Before its definition, we need to highlight some terminologies. 
\begin{definition}[homological critical value]
Let $\mathbb{X}$ be a topological space, and a function $f$ maps $\mathbb{X}$ to $\mathbb{R}$. A homological critical value of $f$ is a real number $a$ such that $\forall \epsilon > 0$ $\exists k \in \mathbb{Z}$ s.t. $H_{k}(f^{-1}([-\infty, a - \epsilon]) \rightarrow H_{k}(f^{-1}(-\infty, a + \epsilon])$ is not an isomorphism.  
\end{definition}
We say a function $f$ is \textit{tame} if it has finitely many homological critical values and $H_{k}(f^{-1}(-\infty, a])$ is finite-dimensional for each homological critical value $a$. Similarly, a persistent module $V_{*}$ is tame if it has a tame function. Indeed, any interval module $I_{*}^{i,j}$ is tame since its homological critical values are precisely $i$ and $j$. Then we obtain the definition of a persistence diagram. 
\begin{definition}[persistence diagram]
The persistence diagram $\mathcal{D}(f) \in \mathbb{R}^{2} \cup {\infty}$ of $f$ is the set of pairs of homological critical points $(a_{i}, a_{j})$, counted with multiplicity $\mu_{i}^{j}$, union all points on $y = x$.  
\end{definition}

\begin{figure}[tbhp]
    \centering
    \includegraphics[width = \linewidth]{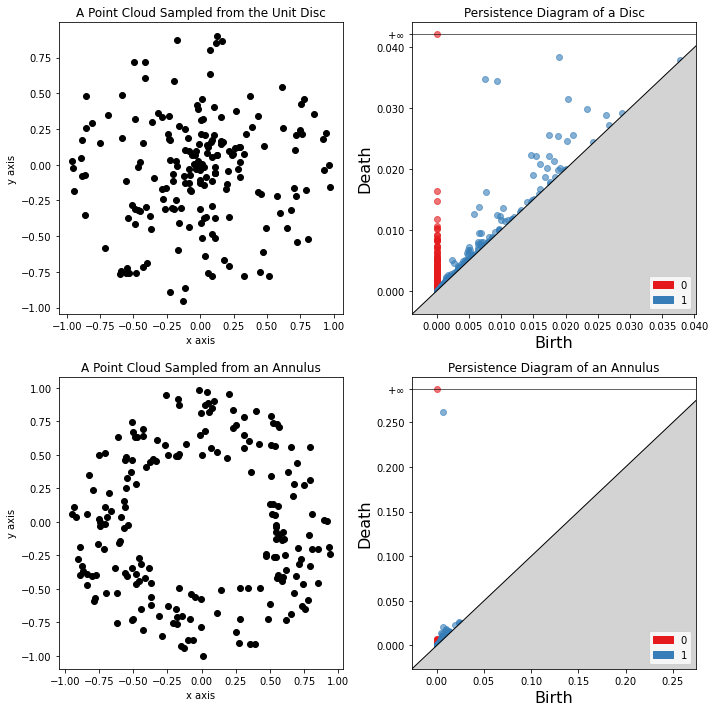}
    \caption{Example point clouds sampled from the unit disc and the annulus $Ann(0.5,1)$ (left column) and their corresponding persistence diagrams (right column) of their Alpha complex filtrations. Red and Blue dots denote persistence of $H_{0}$ and $H_{1}$ homology groups respectively. The red point on the horizontal line marked with $+\infty$ in each diagram represents the immortal connected component.}
    \label{figPD}
\end{figure}

Figure \ref{figPD} shows point clouds consisting of 200 points sampled from the unit disc $\mathbb{D}^{2}$ and the annulus $Ann(0.5,1)$ and their corresponding persistence diagrams. A persistence diagram is a different visualization method of the barcode on the first quadrant by plotting each interval $[b,d]$ as a point $(b,d)$. Since $b \leq d$ always, all points in a persistence diagram locates above the line $y = x$. The persistence diagram of the disc depicts persistences from all dimensions are born at early phases and demise rapidly. However, in the case of the annulus, there exists the 1-dimensional persistence which dies around at $t = 0.25$. Note that these persistence diagrams have different $x$ and $y$ coordinate limits: the inner boundary of the annulus produces distinct topological features. Unlike the disc, the coverings of points near the inner boundary form nerve as a cycle of 1-simplices, which dies out when the whole intersection of the cover becomes nonempty. On the other hand, points are distributed evenly, preventing the formation of such loops. Still, there exists plenty of small loops, but they can be ignored as topological noise. Since their $H_{0}$ persistence is similar, it is the unique topological difference according to the diagrams, and this also matches nicely to the theoretical difference of homology between the two objects.

Here is the definition of the bottleneck distance. 
\begin{definition}[bottleneck distance]
Let $\mathcal{D}$ and $\mathcal{E}$ be persistence diagrams. Let $\eta:$ $\mathcal{D} \rightarrow \mathcal{E}$ be any bijection. Then the bottleneck distance between $\mathcal{D}$ and $\mathcal{E}$ is given by 
\begin{equation}
    d_{B}(\mathcal{D}, \mathcal{G}) = \inf_{\eta}\sup_{\textbf{x} \in \mathcal{D}}\Vert x - \eta(x) \Vert.
\end{equation}
\end{definition}
Now we have one of our main theorems, the stability theorem \cite{cohen2007stability}. 
\begin{theorem}[Stability Theorem]
Let $\mathcal{X}$ be a triangulable topological space with tame functions $f$, $g$. Then the following inequality holds: 
\begin{equation}
    d_{B}(D(f), D(g)) \leq \Vert f - g \Vert_{\infty}.
\end{equation}
\end{theorem}
So, the bottleneck distance between any two persistence diagram is bounded by the $L^{\infty}$ distance between their tame functions. The theorem implies the metric is independent of the noise of the data.

\subsection{Machine Learning with TDA}
\subsubsection{Persistence Landscape}

Since the barcode is a multiset of intervals, it is hard to handle it in machine learning. Persistence landscape \cite{bubenik2015statistical} is a method to vectorize the barcode, making it statistically tractable. 

Let $M$ be a persistent module consisting of the filtration of a complex of the point cloud. For any pair of real numbers $i \leq j$, define the betti number by
\begin{equation}
    \beta^{i,j} = \dim{\img{a_{i \rightarrow j}}}.
\end{equation}
Then we have $\beta^{i,l} \leq \beta^{j,k}$ for any quadruple $i \leq j \leq k \leq l$ since $a_{i \rightarrow l} = a_{k \rightarrow l} \circ a_{j \rightarrow k} \circ a_{i \rightarrow j}$. So for each interval module in the barcorde which is born at $b$ and dies at $d$, define the \textit{rank function} \cite{bubenik2015statistical} $\lambda'(b,d)$ as 
\begin{equation}
    \lambda'(b,d) =
    \begin{cases}
    \beta^{b,d} \hspace{0.5cm} \text{if} \hspace{0.1cm} b \leq d \\
    0 \hspace{0.5cm} \text{otherwise}.
    \end{cases}
\end{equation}
So $\lambda'$ returns the corresponding Betti number only if the input interval is a well-defined interval module in the barcode. Then, define the \textit{rescaled rank function} \cite{bubenik2015statistical} given by 
\begin{equation}
    \lambda(m,h) = 
     \begin{cases}
    \beta^{m-h,m+h} \hspace{0.5cm} \text{if} \hspace{0.1cm} b \leq d \\
    0 \hspace{0.5cm} \text{otherwise}.
    \end{cases}
\end{equation}
where 
\begin{equation}
    m = \frac{b + d}{2}, \hspace{2cm} h = \frac{d - b}{2}.
\end{equation}
Similarly, we have for $0 \leq h_{1} \leq h_{2}$, $\lambda(t, h_{1}) \geq \lambda(t, h_{2})$. Now we have the definition of the persistence landscape \cite{bubenik2015statistical}.

\begin{figure}
    \centering
    \includegraphics[width = \linewidth]{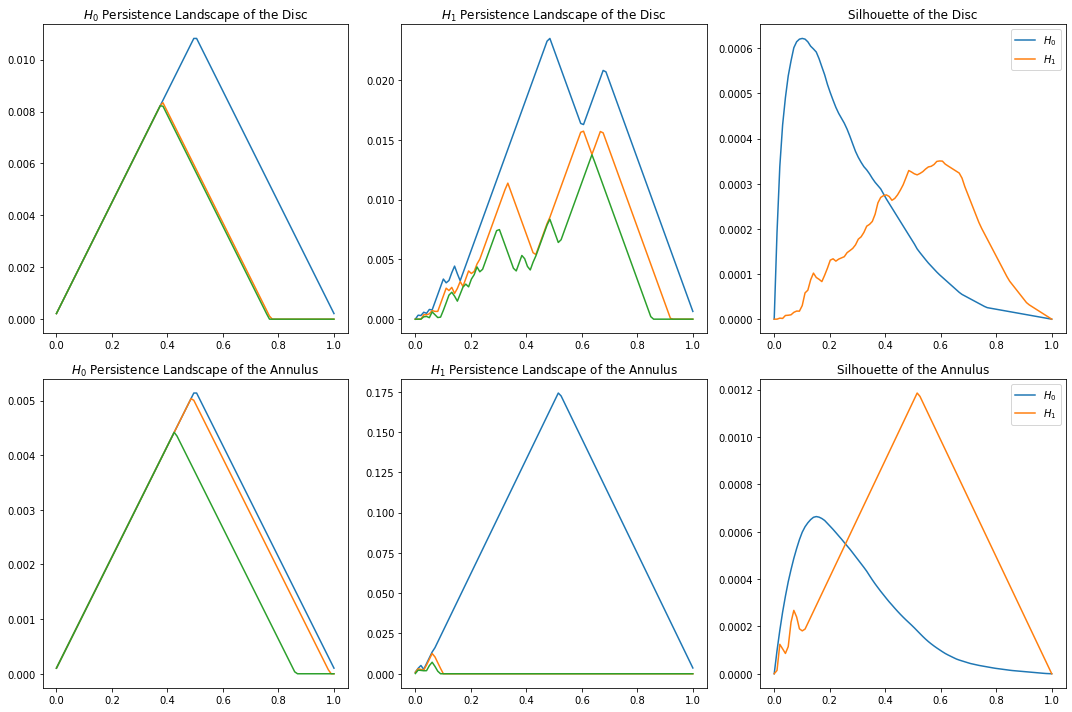}
    \caption{Persistence landscapes of the disc and annulus point clouds in Figure \ref{figPD} and their silhouettes. Three paths (k = 1 (blue), 2 (orange), 3 (green) in equation 25) are sampled with resolution 100 from each persistence diagrams. On the right column, we use the constant weight function to produce the silhouettes.}
    \label{figLandscape}
\end{figure}

\begin{definition}[Persistence Landscape]
The \textit{persistence landscape} is a sequence $\lambda = (\lambda_{k})$ of functions $\lambda_{k}:$ $\mathbb{R} \rightarrow \mathbb{R} \cup \{\infty\}$ where 
\begin{equation}
    \lambda_{k}(t) = \sup{\{m \geq 0 \mid \lambda(t,m) \geq k \}}.
\end{equation}
\end{definition}
Contracting the input domain of each $\lambda_{k}$ to $[0,1]$ makes the function to be a path. Figure \ref{figLandscape} illustrates the top three persistence landscape of the point clouds in Figure \ref{figPD}. $H_{0}$ persistence landscapes do not show the significant difference, but $H_{1}$ persistence landscapes show distinct configurations. $\lambda_{1}$ dominates the others in the annulus but shows more stochastic behaviours in the disc. Instead of simply observing persistence diagrams, vectorized entities transform the topological features into more statistics-friendly objects. Also, we do not need to access all $\lambda_{k}$, sufficient to study only the first few landscapes. 

Analogous to the bottleneck distance, the difference between two persistence landscapes is also measurable by defining its norm. However, since a persistence landscape is a group of numerical vectors, the definition of distance is more natural and statistically tractable.

\begin{definition}[Norms for Persistence Landscapes] \ \\
The $p-$norm of a persistence landscape $\lambda$ is given by
\begin{equation}
    \Vert \lambda \Vert_{p} = \left(\sum_{k = 1}^{\infty}\Vert \lambda_{k} \Vert_{p}\right)^{\frac{1}{p}}.
\end{equation}
\end{definition}
The norm manifests probability space $(\Omega, \mathcal{F}, \mathbb{P})$ with the persistence landscape $\Lambda$ as a random variable embedded in the normed space. So for each $\omega \in \Omega$, $X(\omega)$ is the corresponding persistence data, and $\Lambda(\omega) = \lambda(X(\omega))$ \cite{bubenik2015statistical}. Hence for each persistence $X$, we have a random variable $\Lambda$ as the topological summary statistics. 

So let $X_{1}, \dots X_{n}$ be iid samples and $\Lambda^{1}, \dots \Lambda^{n}$ be the corresponding persistence landscapes. The mean of the landscapes is defined as
\begin{equation}
    \bar{\Lambda}^{n} = \bar{\lambda}^{n}_{k}(t) = \frac{1}{n}\sum_{i = 1}^{n}\lambda_{k}^{i}(t).
\end{equation}
Then we obtain two important theorems for statistical inference applying persistence landscapes.
\begin{theorem}[Central Limit Theorem for Persistence Landscapes]
Let $p \geq 2$. If the both first and second moments of $\Vert \Lambda \Vert$ are finite, then 
\begin{equation}
    ({\bar{\Lambda}^{n} - \mathbb{E}[\Lambda]})/({\sigma / \sqrt{n}}) \rightarrow N(0,1) \hspace{0.2cm} \text{as} \hspace{0.2cm} n \rightarrow \infty
\end{equation}

\end{theorem}
\begin{theorem}[Strong Law of Large Numbers for Persistence Landscapes]

\begin{equation}
    \frac{1}{n}\sum_{i}\Lambda^{i} \xrightarrow{a.s.} \mathbb{E}[\Lambda]
\end{equation}
if and only if $\mathbb{E}[\Lambda] < \infty$.
\end{theorem}
Based on the two theorems, one can perform a statistical test, for example \cite{bubenik2015statistical}. The norm in (26) induces $p-$ landscape distance between two persistence modules. Like the bottleneck distance, it satisfies the stability theorem, but it does not require the tameness condition anymore \cite{bubenik2015statistical}. 

Persistence landscapes give a sequence of path. Instead of multiple paths, a silhouette \cite{chazal2013stochastic} of a persistence diagram returns a weighted average of the diagram. Silhouette compresses the whole diagram into a single path per dimension. 
\begin{definition}[Silhouette]
For each persistence point $p = (\frac{d+b}{2}, \frac{d-b}{2})$, where $b$ and $d$ denote the birth and the death of the point, define a function $\zeta_{p}(t)$ as following:
\begin{equation*}
    \zeta_{p}(t) = 
    \begin{cases}
        t-b \hspace{0.5cm} \text{if} \hspace{0.1cm} t \in [b, \frac{d+b}{2}] \\
        d-t \hspace{0.5cm} \text{if} \hspace{0.1cm} t \in [\frac{d+b}{2},d]\\
        0 \hspace{0.5cm} \text{otherwise}.
    \end{cases}
\end{equation*}
Then, the silhouette $S(t)$ is the weighted average of $\zeta_{p}(t)$:
\begin{equation*}
    S(t) = \frac{\sum_{p}w_{p}\zeta_{p}(t)}{\sum_{p}w_{p}}.
\end{equation*}
\end{definition}
Note that $\lambda_{k}(t) = \text{k}\max_{p}\zeta_{p}(t)$ by definition \cite{chazal2013stochastic}, where $\text{k}\max$ denotes the $k-th$ largest value. Figure \ref{figLandscape} represents constant weight silhouettes of the disc and the annulus. As persistence landscape, $H_{0}$ shapes are similar but the maximum values are different, where $H_{1}$ silhouettes are clearly different. 

\subsubsection{Signature Features}

Even though persistence landscape maps topological features to which statistical learning is applicable, it might contain some artefact caused by choice of the feature map \cite{chevyrev2018persistence}. Signature features prevent this by mapping the paths of persistence landscapes into tensor algebra \cite{chevyrev2018persistence}. They characterize features of the sequence of paths \cite{chevyrev2016primer}. Moreover, granted by its tensorized structure, signature transform allows faster computation in both CPUs and GPUs, which is crucial for efficient statistical learning \cite{kidger2020signatory}.

\begin{definition}[Signature]\ \\
The signature of a path $X = (X_t^{1}, \dots X_{t}^{n}):$ $[a, b] \rightarrow \mathbb{R}^{n}$ is a collection of integrals of X such that 
\begin{equation*}
    S(X)_{a,b} = (S(X)_{a,b}^{0},S(X)_{a,b}^{1},\dots,S(X)_{a,b}^{n},S(X)_{a,b}^{1,1}\dots S(X)_{a,b}^{1,2},\dots)
\end{equation*}
with $S(X)_{a,b}^{0} - 0$ and 
\begin{equation}
    S(X)_{a,t}^{i_{1},\dots,i_{k}} =  \int_{a < t_{k} < t} \cdots \int_{a < t_{1} < t_{2}} dX_{t_{1}}^{i_{1}} \cdots  dX_{t_{i}}^{i_{k}}. 
\end{equation}
\end{definition}
The $k-th$ level signature is the collection of $S(X)^{i_{i}, \dots i_{k}}$ such that $1 \leq i_{i}, \dots i_{k} \leq k$. Hence, a $k-th$ level signature is composed of $n^{k}$ values. One important property of the signature is its independence of time reparametrization. Suppose $X,Y$ are path with domain $[a,b]$, and let $\psi: [a,b] \rightarrow [a,b]$. Let $\Tilde{X}_{t} = X_{\psi(t)}$ and $\Tilde{Y}_{t} = Y_{\psi(t)}$ defined on the same domain. Then we have 
\begin{equation*}
    \Dot{\Tilde{X}}_{t} = \Dot{X}_{\psi(t)}\Dot{\psi}(t) 
\end{equation*}
which leads to 
\begin{equation}
    \int_{a}^{b}\Tilde{Y}_{t}d\Tilde{X}_{t} = \int_{a}^{b}Y_{\psi(t)}\Dot{X}_{\psi(t)}\Dot{\psi}(t)dt =
    \int_{a}^{b}Y_{u}du 
\end{equation}
by substituting $u = \psi(t)$ \cite{chevyrev2016primer}. Therefore, since the signature is a nested integral in (31), $S(\Tilde(X))_{a,b}^{i_{1}, \dots i_{k}} = S(X)_{a,b}^{i_{1}, \dots i_{k}}$ for all $i_{m} \in \{1,\dots,n\}$. 

Another important property is the shuffle identity \cite{chevyrev2016primer,ree1958lie}. A $(k,m)-shuffle$ of the set $\{1,\dots,k+m\}$ is a permutation $\sigma$ on the set such that $\sigma^{-1}(1) < \dots < \sigma^{-1}(k)$ and $\sigma^{-1}(k+1)< \dots < \sigma^{-1}(k+m)$. $Sh(k,m)$ indicates the set of all (k,m)-shuffles.  Let $I = (i_{1}, \dots , i_{k})$ and $J = (j_{1}, \dots j_{m})$ be two multi indexes, $i_{1}, \dots, i_{k}, j_{1}, \dots j_{m} \in \{1, \dots d\}$.  
\begin{definition}[Shuffle product] The shuffle product $I\#J$ of $I$ and $J$ is the set of multi-indexes such that 
\begin{equation}
    I\hspace{0.1cm} \# \hspace{0.1cm} J = \{(r_{\sigma(1)}, \dots r_{\sigma(k+m)} \mid \sigma \in Sh(k,m) \}.
\end{equation}
\end{definition}
\begin{theorem}[Shuffle identity \cite{chevyrev2016primer}]
For a path $X$: $[a,b] \rightarrow \mathbb{R}^{n}$ and multi-indexes $I = (i_{1}, \dots i_{k})$, $J = (j_{1}, \dots , j_{m})$ whose entries are from $\{1,\dots,n\}$, 
\begin{equation}
    S(X)^{I}_{a,b}S(X)^{J}_{a,b} = \sum_{K \in I \hspace{0.1cm} \# \hspace{0.1cm} J}S(X)^{K}_{a,b}.
\end{equation}
\end{theorem}
This enables product of the values across the level. For instance, when $n = 2$, \\ $S(X)_{a,b}^{1}S(X)_{a,b}^{2} = S(X)_{a,b}^{1,2} + S(X)_{a,b}^{2,1}$ and $S(X)_{a,b}^{1}S(X)_{a,b}^{1,2} = S(X)_{a,b}^{1,1,2} + S(X)_{a,b}^{1,2,1}$. 
The shuffling identity hence simplifies the arbitrary product into linear combination of elements from higher channel.

So, if we take persistence landscape as input, $t_{1}, \dots, t_{k}$ corresponds to the points where the landscape values are sampled. For instance, in Figure \ref{figLandscape}, sequence of $t_{i}$'s is the linspace(0,1,20). Features extracted through the signature transform is widely used in machine learning, especially models concerning time series data prominent in finance \cite{gyurko2013extracting}.
\section{Results}

\subsection{Data Description}

The data consists of three cell types: Astrocyte, Microglia, and Neuron. Astrocyte and Microglia are glial cells in the brain which support neuronal activities. Initially, iPSC (induced pluripotent stem cell) lines were acquired from StemBANCC and nurtured. Differentiation of stem cells directed to cortical neurons reaches astrocyte or neuron \cite{liu2013direct}. Microglia nurture in monocyte plants after the embroid state\cite{haenseler2017highly}. 

Before the stain treatment, cells of each cell types are plated on 384 well plates, coated and incubated. They are thawed, resuspended into corresponding media, seeded into wells and cultured. Cells are imaged in four stains plus Brightfield, and table \ref{tab1} shows the name of compound used for staining. MitoTracker Deep Red is added to wells at a concentration of 500 nM followed by PFA for cell fixation. Afterwards, remaining Concanavalin A Alexa488 conjugate, Phalloidin, and Hoescht 33342 are added. After resting cells absorb treatments, they are washed twice with PBS (Phosphate-buffered saline). 
 
\begin{figure}[tbhp]
    \centering
    \includegraphics[width = \linewidth]{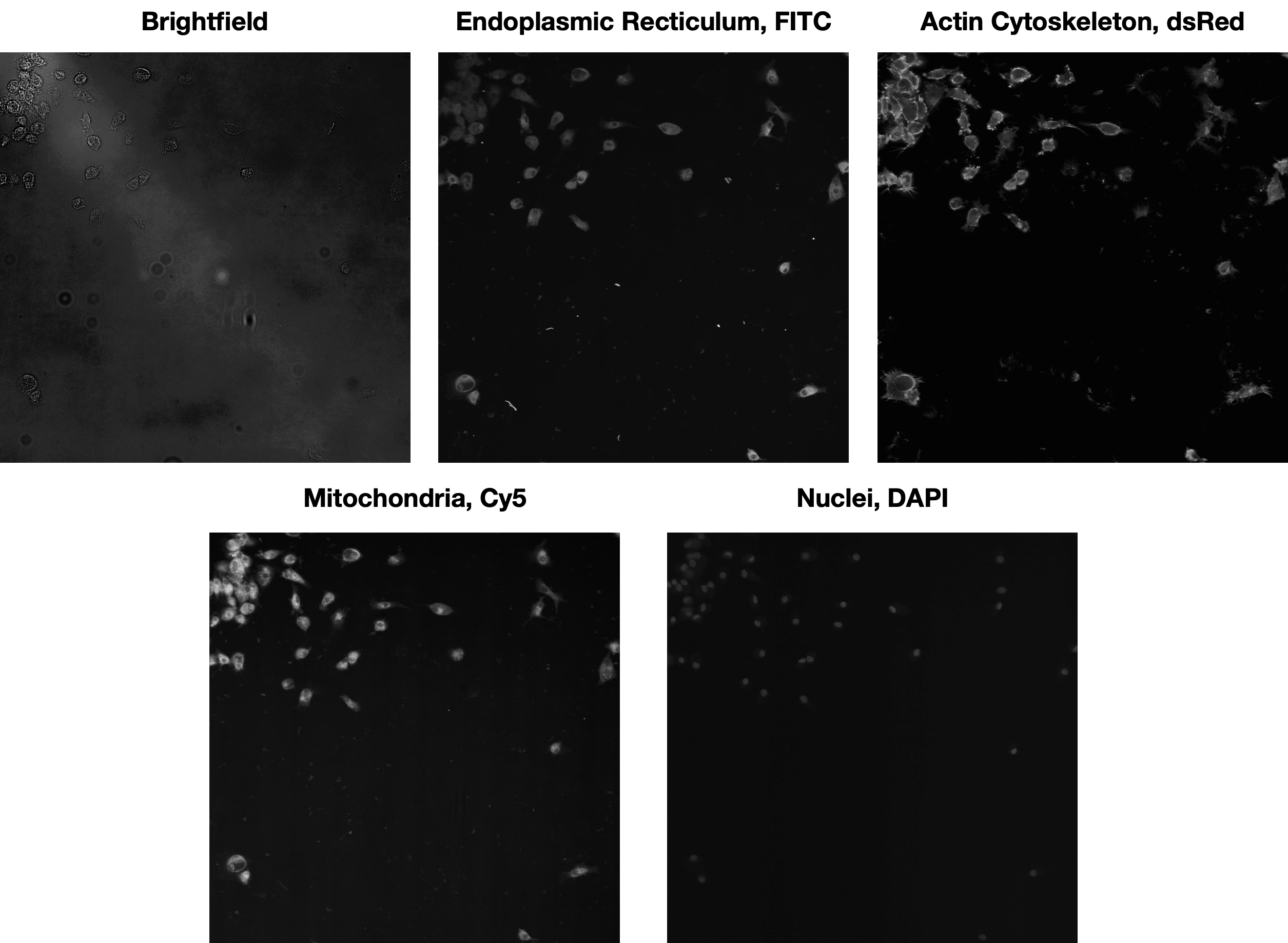}
    \caption{Four stains and a brightfield image of Microglia from the C - 10 well of the first plate of the cell line 803. Given Images are enhanced from the raw data for clearer display.}
    \label{figMgl}
\end{figure}

    \begin{table}
        \centering
        \begin{tabular}{|l | c |c |}
        \hline
        Name & Marker & Image channel  \\
        \hline
         MitoTracker & Mitochondria & Cy5 \\
        \hline
        Concanavalin A & Endoplasmic Reticulum  & FITC \\
        \hline
        Phalloidin & Actin cytoskeleton & dsRed \\
        \hline
        Hoescht 33342 & Nuclei & DAPI \\
        \hline
        BF & None (bright field image) & TL-Brightfield \\
        \hline
        
    \end{tabular} 
\caption{Staining compounds assigned to their target organelles and corresponding image channels.}
        \label{tab1}
    \end{table}
There are 19,200 grayscale tiff images of size $2048 \times 2048$ per plate and five plates per cell line. Images are collected on the IN Cell Analyzer 6000 with a 20X objective, followed by image enhancement through CellProfiler and sequence of Cell Painting analysis. Cell lines 803 and 808 corresponds to AD patients with PSEN1 mutation \cite{fortea2010increased}, while 840 and 856 to people with no AD. So we can develop a semantic segmentation model which can capture cell organelles from each cell type. Figure \ref{figMgl} shows an example set of Microglia images.

\subsection{Deep Learning Simulations}
\subsubsection{Multiclass Semantic Segmentation}
\begin{figure}[tbhp]
    \centering
    \includegraphics[width = 0.8\linewidth]{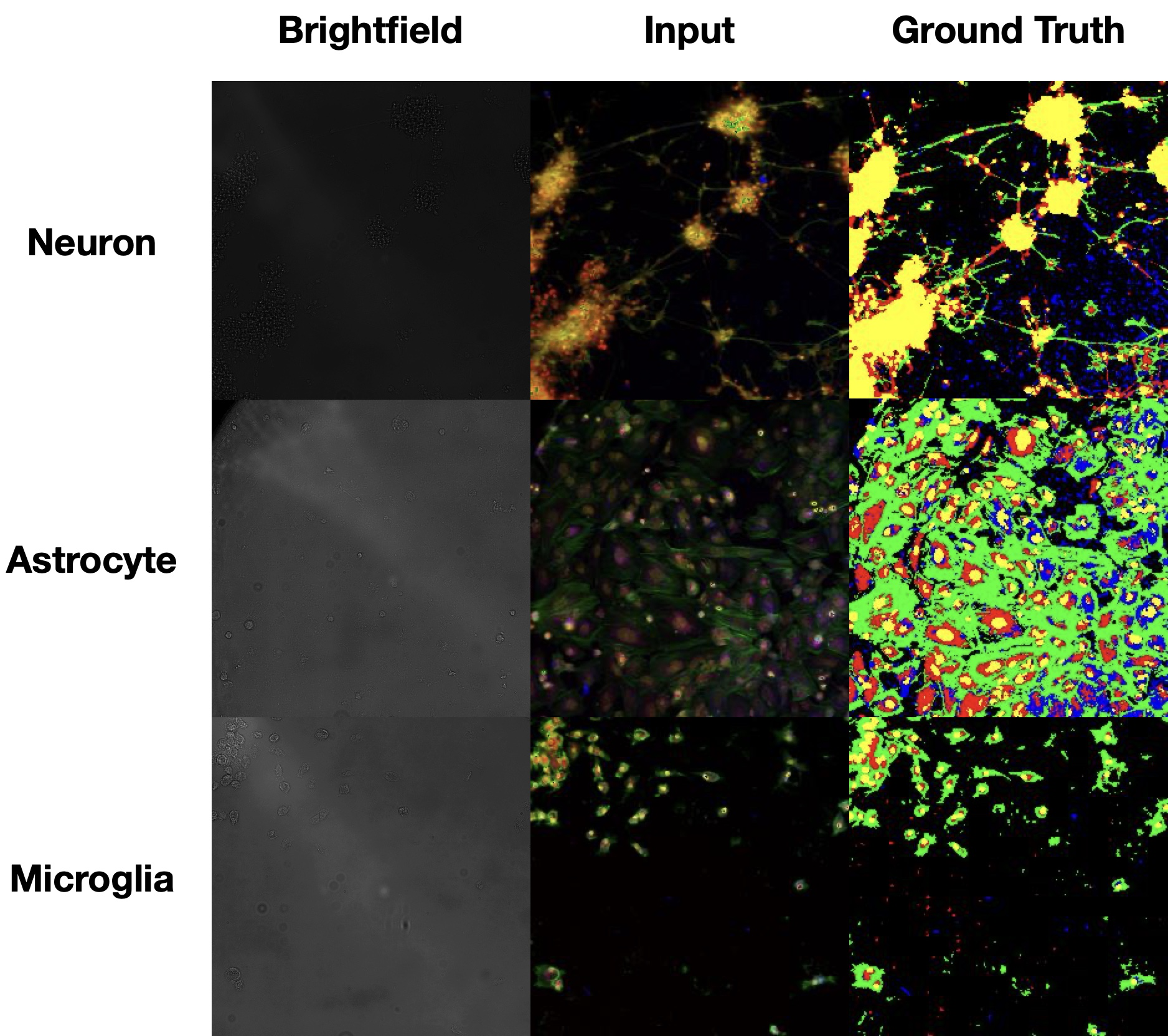}
    \caption{Brightfield images and their corresponding ground truth and stacked inputs.}
    \label{SN_input}
\end{figure}

We build a semantic segmentation model to partition all four cell organelles taking stacked RGB images as inputs. We do not choose the brightfield as input since its features are too insufficient to be unravelled. We have five images from each cell well. Neglecting the brightfield images, we construct a stacked RGB image by accumulating stains of mitochondria, cytoskeleton, and ER, preserving the order followed by summing nuclei image to the first and the second channel, making it yellow (red plus green is yellow) in the visualization of the input. Figure \ref{SN_input} shows how the target label of our input looks like. Images are resided to $128 \times 128$ to lighten computation and inputs are normalized prior to the training.

We need the 'ground truth' to train a semantic segmentation model. We threshold the stained images to produce the labels. Pixels are normalized and fitted into $[0,255] \cap \mathbb{Z}$. Label and input pairs of all three cell types are shown in Figure \ref{SN_input}. Each ground truth image is transformed to a 1-channel image whose pixel values are in $\{0,1,2,3,4\}$. The background pixel has value 0, where the assignment follows the order in table $\ref{tab1}$. 

We admit cross-entropy loss for the risk minimization. Remark the object of image segmentation is pixel-wise classification. So if a pixel has an output
\\
$[p_{0},p_{1}, \dots, p_{n}]$ whose corresponding true label is $i$, the model would like to know how much $[p_{0},p_{1}, \dots, p_{n-1}]$ and $\delta_{i,0}, \delta_{i,1}, \dots, \delta_{i,i}, \dots, \delta_{i,n-1}$ are different. So it is equivalent to compute the loss as 
\begin{equation*}
    L(x, i) = -\log{\left(\frac{\exp{x[i]}}{\sum_{j}\exp{x[j]}}\right)}
\end{equation*}
if there are $n$ classes including the background in total. 

\begin{figure}
    \centering
    \includegraphics[width = \linewidth]{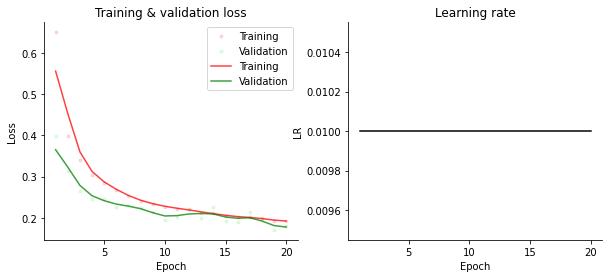}
    \caption{Training and Validation losses over 20 epochs (left) and learning rate propagation (right). Learning rate is fixed over the training.}
    \label{figUnet experiment}
\end{figure}

\begin{figure}
    \centering
    \includegraphics[width = \linewidth]{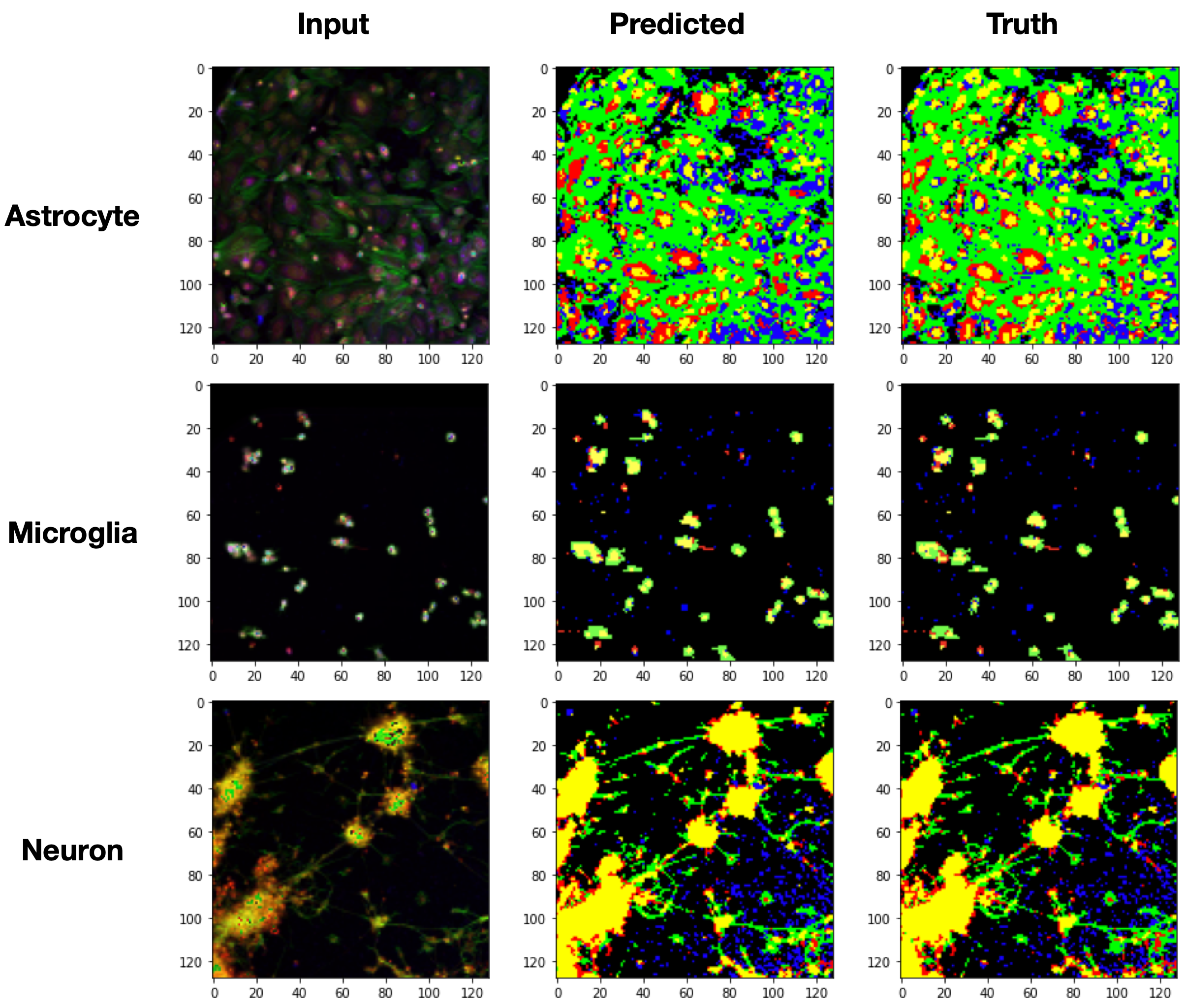}
    \caption{Input stacked RGB images, model prediction and ground truth of an example in validation set from each cell type. }
    \label{figUnetoutput}
\end{figure}

Figure \ref{figUnet experiment} shows the training and validation losses of the Unet model. We omitted the scheduler and selected SGD for the optimization method with learning rate = 0.01, running 20 epochs. The training set consists of all cell types, 1000 images for each cell type. Overfitting does not occur during the training. Figure \ref{figUnetoutput} shows the sample test images and their predictions. 

\begin{figure}
    \centering
    \includegraphics[width = \linewidth]{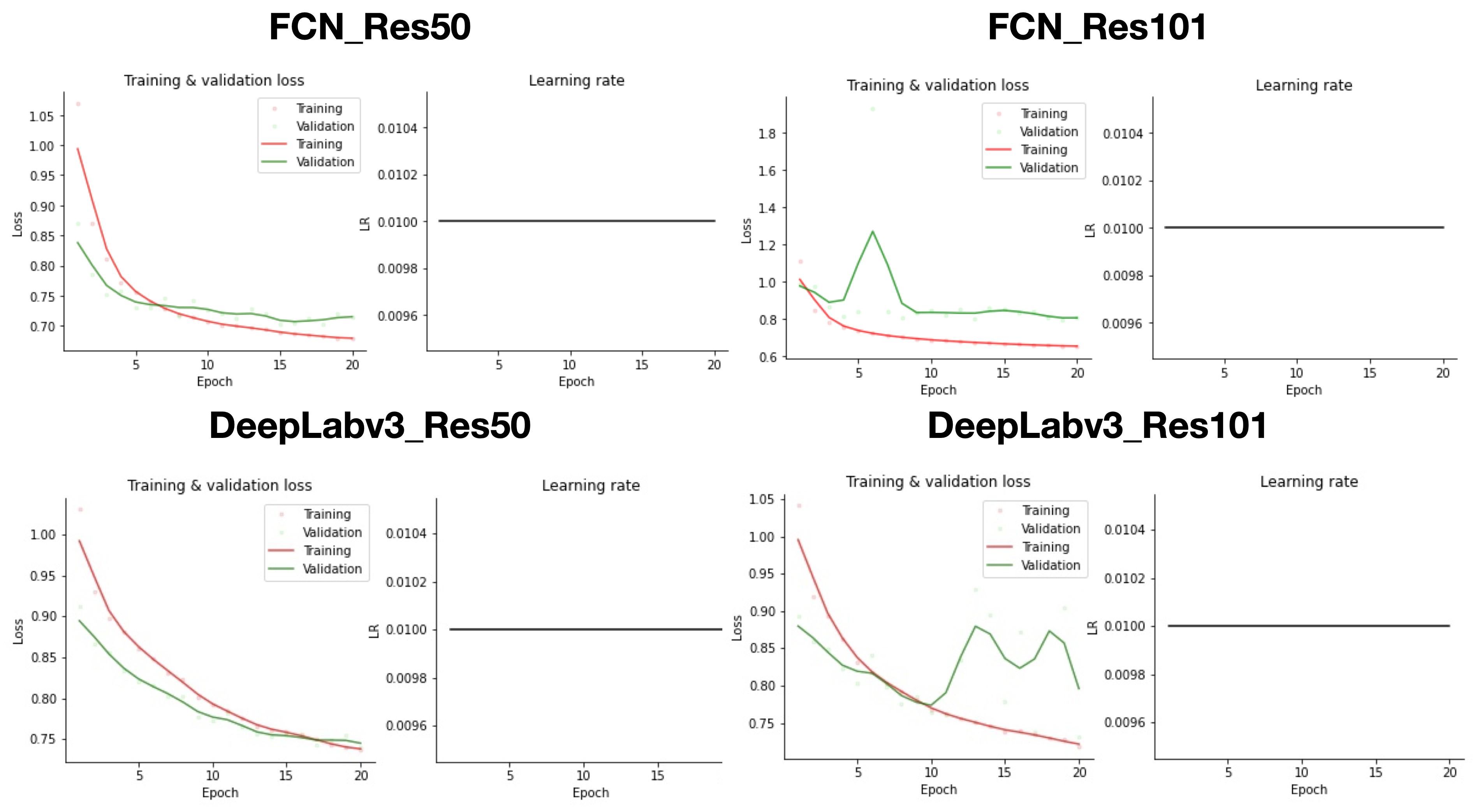}
    \caption{Training and validation losses plot of FCN and DeepLabv3 with Resnet-50 and Resnet-101 backbones.}
    \label{figFCNDL}
\end{figure}

However, performances of FCN and DeepLabv3 are inferior to Unet in terms of training and validation losses. Figure \ref{figFCNDL} shows training and validation losses of FCN and Deeplabv3 models with two different encoding backbones, Resnet-50 and Resnet-101 where the other conditions are remaining invariant. We expect this is due to the coarse downsampling process in both models, where the Unet has multiple decoding procedures. Since our cell image segmentation model has a small number of classes but sharp boundaries, only a few decoding layers are insufficient for detailed segmentation. 

    \begin{table}
        \centering
        \begin{tabular}{|c | c| c| c|}
        \hline
        Loss type & FCN & Deeplabv3 & Unet \\
        \hline
        Training loss & \makecell{0.6791 (Res50) \\ 0.6526 (Res101)} & \makecell{0.7365 (Res50) \\ 0.7180 (Res101)} & \textbf{0.1913} \\
        \hline
        Validation loss & \makecell{0.7138 (Res50) \\ 0.8092 (Res101)} & \makecell{0.7400 (Res50) \\ 0.7315 (Res101)} & \textbf{0.1791} \\
        \hline
    \end{tabular} 
\caption{Training and validation losses of five models in the multi-class semantic segmentation at the last epoch.}
        \label{tab2}
    \end{table}

Until now, the models are trained with images of all three cell types, but this leads to a question if the model is transferable: showing high accuracy in a different class of data. Suppose we have a model only trained with astrocyte data. Will the model be consistent in segmenting microglia or neuron images? 

\begin{figure}
    \centering
    \includegraphics[width = \linewidth]{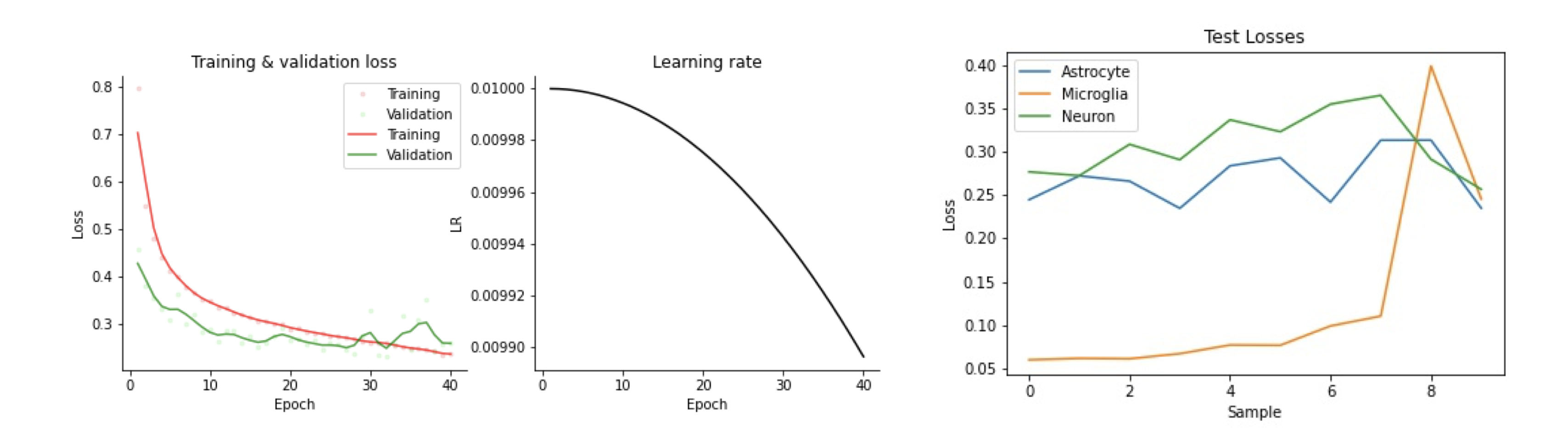}
    \caption{Left: training and validation losses of the transfer learning model trained only with astrocyte data. The learning rate plot against the number of epochs passed is illustrated on the right. Training and validation losses are 0.2375 and 0.2620 after the last epoch. Right: Test losses of astrocyte, microglia, and neuron images. The mean test losses along all batches are 0.2696 (Astrocyte), 0.1257 (Microglia), and 0.3075 (Neuron).}
    \label{segtransfer}
\end{figure}

Figure \ref{segtransfer} shows the model for the transfer learning task. We train the model with 3000 Astrocyte images from cell line 803, sticking on the Unet structure due to its better performance in our regime. Additionally, we include the cosine annealing scheduler after each validation step in every epoch. After running 40 epochs, we tested the model by the cross-entropy loss using 100 images from each cell type. After partitioning each test class into ten batches, we compute the average test loss per batch. 

The plot on the right of Figure \ref{segtransfer} shows all three test losses do not show a high discrepancy to the validation loss of the model. Especially, test losses of the microglia dataset are significantly lower than the others since microglia images consist of a higher proportion of backgrounds (see Figure \ref{figUnetoutput}). 
\subsubsection{Augmented Microscopy}
Unfortunately, the multi-class segmentation is impractical in biology research; we are re-stacking images already segmented by the cell painting. Still, it shows the Unet is the most appropriate for our approach. 

However, segmenting brightfield images is demanding in real-world research indeed. In Figure \ref{SN_input}, brightfield images do not show explicit features, but can a deep learning model figure out their latent features? If there is such an \textit{in-silico} approach, it will accelerate the research procedure with high accuracy. 

Figure 19 displays the result of augmented microscopy experiment. We have 3000 images of astrocytes, and run them for 30 epochs with batch size 16. We use parallel computing with 3 GPU’s internally supported by Pytorch. Training is not different to the previous simulations, but we exploit Intersection-over-Union (IoU) score, or Jaccard’s index, to quantify the performance of the model. For each class $i$, we count the number of overlapping pixels of value $i$ (say $i$-pixel) in the both divided by number of  $I$-pixels occurring at least one of the truth of the prediction. For instance, in Figure \ref{fig20}, IoU of the class 1 is $Yellow / (Yellow + Red_{truth} + Red_{pred}) = 6 / (6 + 3 + 3) = 0.5$, and that of the class 0 is $13 / (13 + 3 + 3) ~ 0.68$. The total IoU is computed by averaging all class-wise IoU’s, so $(0.5 + 0.68) / 2 = 0.59$.

\begin{figure}[tbhp]
    \centering
    \includegraphics[width = \linewidth]{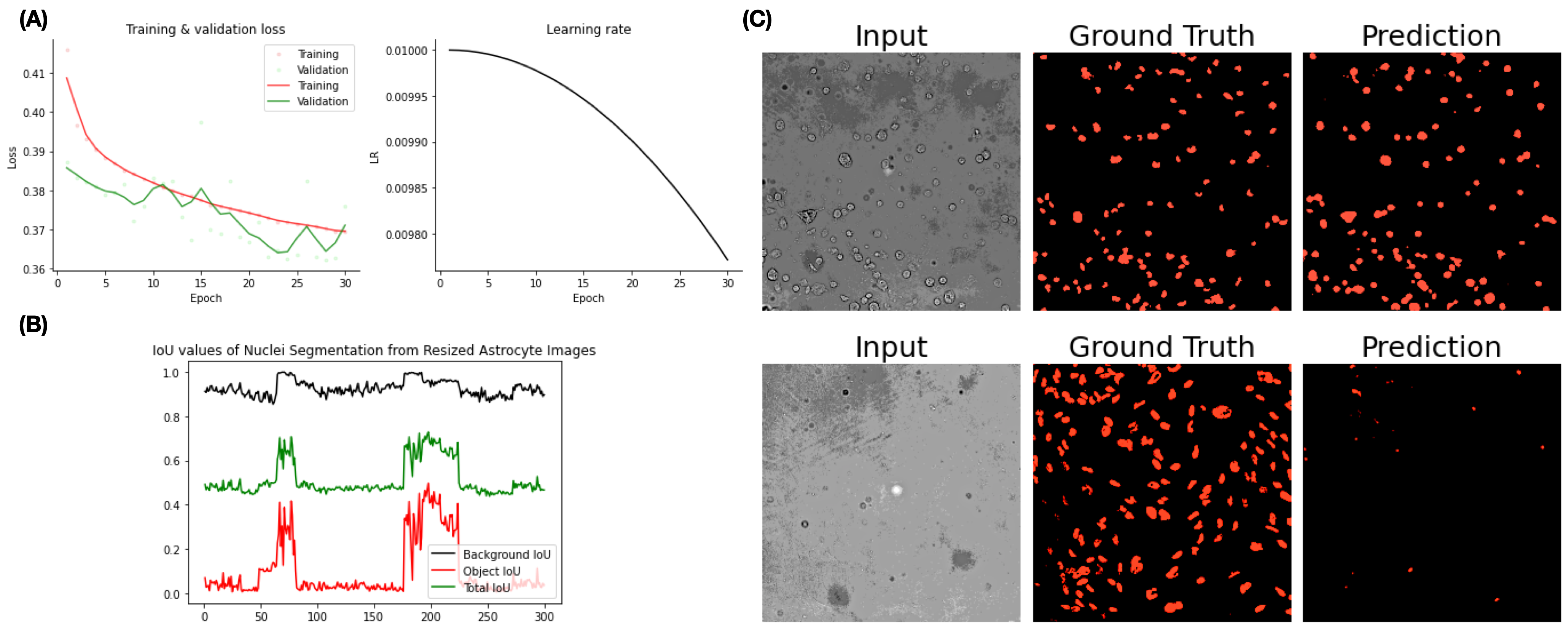}
    \caption{Training procedure and test results of Unet based augmented microscopy model. (A) Training and validation losses over 30 epochs. Learning rates are plotted on the left, decreasing due to the cosine annealing scheduler. (B) Background, object, and total IoU (Intersection-over-Union) score of the test image set consisting of 200 images. (C) (Input, Prediction, Truth) triples in the test set with the largest (top) object IoU and the smallest (bottom) object IoU.}
    \label{fig19}
\end{figure}

\begin{figure}[tbhp]
    \centering
    \includegraphics[width = 0.7\linewidth]{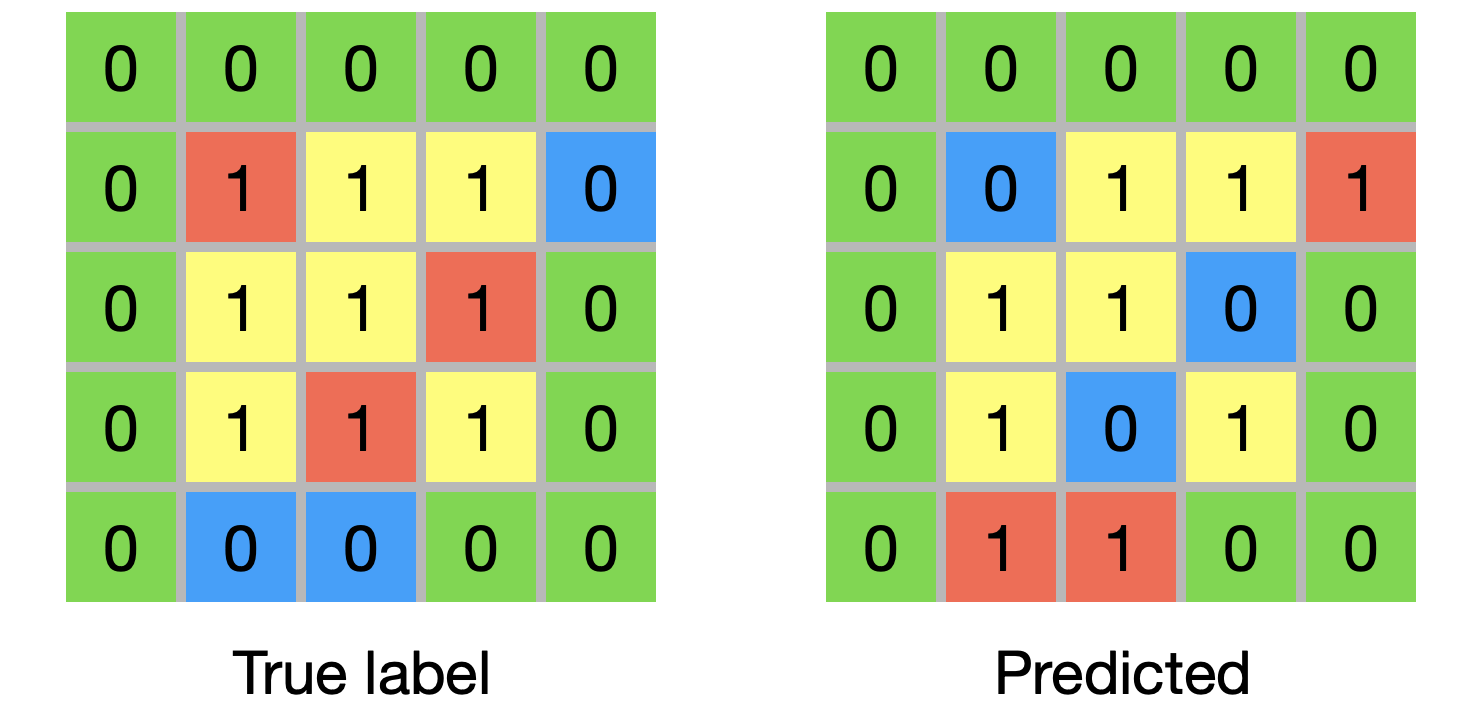}
    \caption{An example of finding IoU score.}
    \label{fig20}
\end{figure}

In Figure \ref{fig19}, images are resized before the training but now to $512 \times 512$ for higher resolution. In terms of IoU, the performance is not optimal since object IoU is mostly close to 0. However, for some training examples, the object IoU bursts up to 0.45. Surprisingly, comparing the input images of maximum object IoU against minimum object IoU, nuclei are much more explicitly displayed in the brightfield image in the former but not in the latter. It is because brightfield images can be noisy, unlike carefully harvested fluorescent channel images. This result provides whatever the input is, the model learns the feature of the input images regardless of the ground truth.

\begin{figure}[tbhp]
    \centering
    \includegraphics[width = \linewidth]{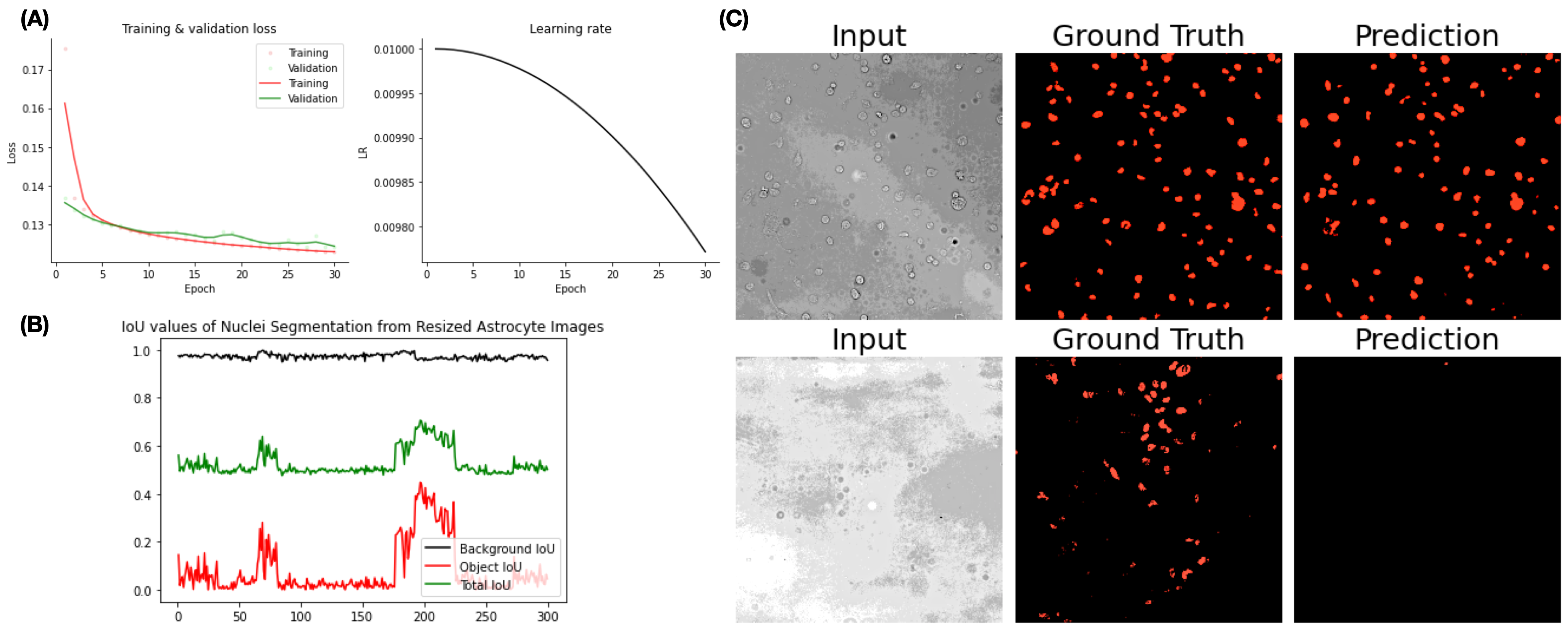}
    \caption{Everything is the same with Figure \ref{fig19} but the ground truths are sampled through thresholding all pixels which have top 5\% strength. }
    \label{fig21}
\end{figure}

Now the ground truth producing protocol changes to thresholding pixels with top 5\% strength. Training and validation losses get smaller as Figure \ref{fig21} (C) shows fewer mismatching pixels of which has smaller IoU. However, there is no considerable difference in total IoU; the latter has a higher maximum object IoU but compromised by lower average background IoU. 

\begin{figure}[tbhp]
    \centering
    \includegraphics[width = \linewidth]{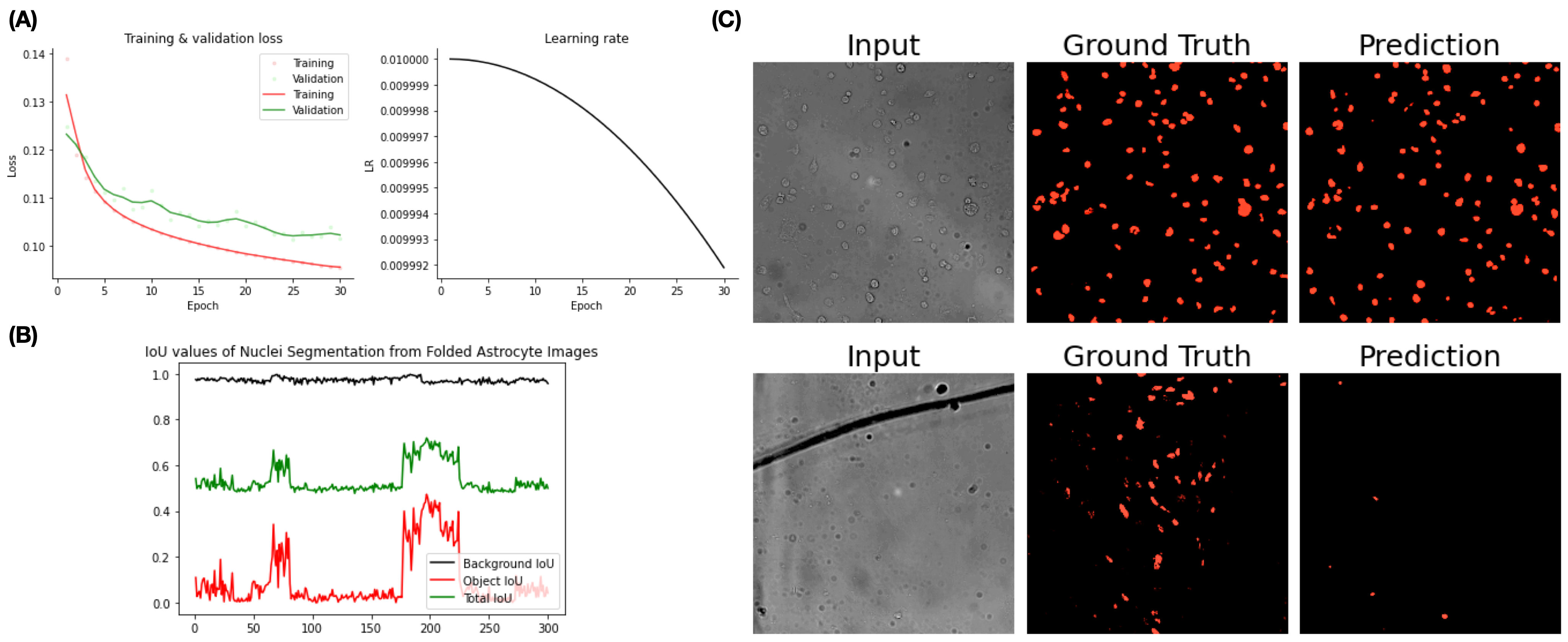}
    \caption{Results from the folded dataset. IoU’s are computed after aggregating the sliced images into the original size. }
    \label{fig22}
\end{figure}

Resizing might lose vital information contained in every pixel. We also examine folded images, where inputs produced by partitioning a raw $2048 \times 2048$ image to 16 $512 \times 512$ images. We folded 2000 brightfield images, so the dataset has 32000 images. In Figure \ref{fig22}, though it gives smaller training losses, its performance is similar to the past models for maximum object IoU and average total IoU. Therefore, we can conclude pre-transformation procedures are independent of the model performance. We will maintain the last option for the remaining simulations. 

As the multi-class segmentation, we test the model transferablility. Table \ref{tab3} and \ref{tab4} represent IoU's of three cell type test data of models trained only with astrocyte data and the all cell types respectively. Comparing the third row, performance for testing astrocyte is decreased but increased or remained constant for other cell types. Big rise of Microglia IoU is notable. Considering number of astrocyte training set decreased, we confirm training the augmented microscopy model with all three data shows improved overall performance.
\begin{table}[tbhp]
        \centering
        \begin{tabular}{|c | c| c| c|}
        \hline
        IoU & Astrocyte & Microglia & Neuron \\
        \hline
        Average Object IoU & 0.0964 & 0.1181 & 0.4178 \\
        \hline
        Average Background IoU & 0.9722 & 0.9742 & 0.9469 \\
        \hline
        Average Total IoU & 0.5343 & 0.5462 & 0.6823\\
        \hline
        Maximum Object IoU & 0.4731 & 0.3992 & 0.7094\\
        \hline
    \end{tabular} 
\caption{IoU scores from testing the model trained with only astrocyte images.}
    \label{tab3}
\end{table}

\begin{table}[tbhp]
        \centering
        \begin{tabular}{|c | c| c| c|}
        \hline
        IoU & Astrocyte & Microglia & Neuron \\
        \hline
        Average Object IoU & 0.0614 & 0.4745 & 0.4038 \\
        \hline
        Average Background IoU & 0.9719 & 0.9839 & 0.9610 \\
        \hline
        Average Total IoU & 0.5167 & 0.7292 & 0.6824\\
        \hline
        Maximum Object IoU & 0.3908 & 0.7891 & 0.7332\\
        \hline
    \end{tabular} 
\caption{IoU scores of the model trained with all three types of cell images, 1000 from each.}
    \label{tab4}
\end{table}

\subsection{Topological Data Analysis Simulations}

We are going to classify fluorescent nuclei images by its cell type only using topological features, naming the topological transformer of the data. As the toy simulation in the section 3.2.1, it requires to transform an input raw image into a point cloud. Here we use two methods. One is that we resize a $2048 \times 2048$ image to a $64 \times 64$ image and regard every nonzero pixel as a point in the plane. Although the image is robustly deformed, information relating to cell size and distance between cells are well-preserved by constrained proportion. The other approach is first thresholding all nonzero pixels as we produced the ground truth in the semantic segmentation experiment, and draw contours of each target area. Then, using OpenCV library, we plot the centre of each contour. 
\begin{figure}
    \centering
    \includegraphics[width = \linewidth]{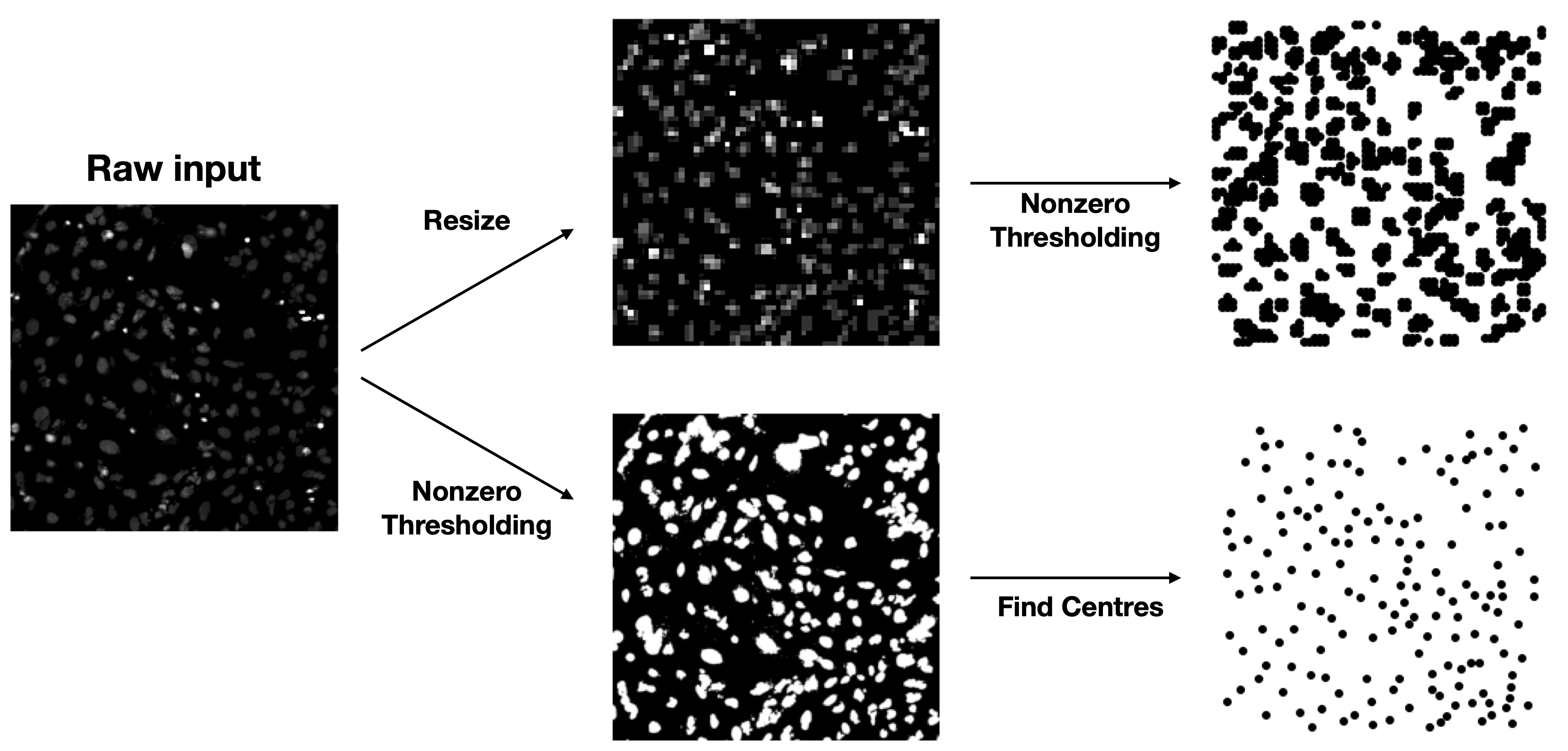}
    \caption{Resizing (top) and Contour (bottom) methods for point cloud generation.}
    \label{figprep}
\end{figure}

We are going to examine two different topological transformers: using the silhouette or the signature. Primarily, we compute the persistence diagram of the alpha complex of a point cloud. Silhouette transformer computes the silhouette of the constant weight per dimension. Here we fix the maximum dimension to 1 and the resolution of the silhouette to 200. Stacking the sequence along the channel gives $2 \times 200$ data. The signature transformer generates the signature from the top five persistence landscapes of each dimension. It accumulates signatures up to the third level. So the length of each signature is $5^{0} + 5^{1} + 5^{2} + 5^{3} = 156$, but we omit the first term since it is always 1 and could dominate the remaining sequence. So, we have $2 \times 155$ arrays as input. 

We first perform classification using three statistical machine learning classifiers: Support Vector Machine (SVM), XGBoost \cite{Chen_2016}, and Light Gradient Boosting Machine (LGBM) \cite{NIPS2017_6449f44a}. Both XGBoost and LGBM root on the Gradient Boosting Decision Tree algorithm, but LGBM accelerate the training by exclusive feature bundling plus gradient-based one-side sampling \cite{NIPS2017_6449f44a}. 

Table \ref{tab5} shows the resizing results in better performance for all three classifiers. We infer this as resizing eliminates more noises. For resizing, XGBoost win by a margin, but LGBM shows higher accuracy in the contour method but with a larger difference. So, we conclude LGBM outperforms the other classifiers for all four topological transformers.
\begin{table}[tbhp]
        \centering
        \begin{tabular}{|c | c| c| c|}
        \hline
        Transformer & SVM & XGBoost & LGBM\\
        \hline
        Resize\_silhouette & 78.8\% & \textbf{86.9\%} & 86.6\%  \\
        \hline
        Resize\_signature & 75.6\% & \textbf{80.8\%} & 80.1\% \\
        \hline
        Contour\_silhouette & 58.7\% & 70.4\% & \textbf{71.2\%} \\
        \hline
        Contour\_signature & 57.8\% & 63.6\% & \textbf{65.1\%} \\
        \hline 
    \end{tabular} 
\caption{Test accuracy of the three classifiers toward each topological transformer. Each training set consists of 2000 images per cell type, and training and test set are split into ration 8:2.}
    \label{tab5}
\end{table}

Beyond the statistical approach, we build a TDA-CNN to classify topological features using a simple 1-dimensional convolutional neural network. The CNN we use consists of four convolutional layers followed by two fully connected layers. ReLU activation is present after all layers except the output, and we apply batch-normalization after each convolutional layer. Each convolutional layer doubles the number of input channels with kernel size = 4, stride 2, and padding 1. Table \ref{tab6} shows neither TDA-CNN outperforms the optimal classifier in Table \ref{tab5}. Also, it is the contour\_silhouette method that shows the highest accuracy along with a neural network. Still, all the topological methods fall behind a conventional neural network of an identical structure to TDA-CNN except discharging 2D convolutions.  

\begin{table}[tbhp]
    \centering
    \begin{tabular}{|c|c|}
        \hline
         Transformer & Test Accuracy \\
         \hline
         Resize\_silhouette & 82.8\% \\
         \hline
         Resize\_signature & 77.6\% \\ 
         \hline
         Contour\_silhouette & 84.2\% \\ 
         \hline
         None & 93.8\% \\
         \hline
    \end{tabular}
    \caption{Test accuracy of TDA-CNN with different transformers and a normal CNN without topological preprocessing.}
    \label{tab6}
\end{table}

Table \ref{tab6} displays utilizing only topological features is insufficient for cell image classification. Figure \ref{SN_input} illustrates that astrocyte images contain many small cells stationed densely but sparser in microglia images. On the other hand, neuron images do not contain cells as many as the others, plus some of them are much larger. Such information manifests into the persistence of the image point cloud, but the topological transformers neglect other vital information like the shape of each cell. 

\section{Conclusions}

We confirm Unet is most suitable for our multi-class semantic segmentation task, superior to FCN and DeepLabv3. Also, the transferability of the multi-class semantic segmentation implies the model learns hidden features only based on pixel data, regardless of the cell type. Furthermore, we present the performance of the augment microscopy is independent of the choice of ground truth producing protocol or data scaling. Finally, a model ignores turbulence in the ground truths and successfully adapts necessary latent features. Even though the transferability of the augmented microscopy model is unfavourable, further research in transfer learning of cell image segmentation is promising. In topological data analysis experiments, we found Resize\_silhouette transformer show the best performance for all three SVM, XGBoost, and LGBM classifiers. The Contour\_silhouette is the most suitable for classification involving a neural network, even though none of the TDA-CNN models surpasses the model excluding a topological transformer.   

To improve the results, We first require more refined ground truths: naive thresholding might be unfit in general. A different segmentation model is considerable since Unet is quite old and more modern models are being developed. Moreover, the models are not guaranteed to perform equally to data collected with different methods. Therefore, we should examine the model if a different cell painting or image-collecting method prohibits the transfer of the model. Finally, An image containing numerous cells impairs detailed segmentation. Using high-resolution data comprise a few cells can make segmentation and augmented microscopy ameliorated. 

Our research has lots of potentials for further applications. Augmented microscopy can ease experiment procedure by only examining brightfield images. Also, extension to multi-class augmented microscopy promises massive versatility in biomedical research. We also need to focus on which topological transformer reforms an image into a 1-dimensional sequence while preserving significant geometric features. The visual transformer \cite{dosovitskiy2020image} also converts images into a sequence by the famous transformer method in natural language processing. Transformation into 1D data enables us to examine the images in recurrent neural networks. As the visual transformer reduces computational costs, we expect the topological transformer facilitates it similarly. If the topological features of the input data are explicit, then TDA-CNN can outperform conventional CNN with lower computational cost. For example, suppose A$\beta$ influences cell configuration by scattering cells. If a treatment suppresses the A$\beta$ plagues, then alteration of the topology is recorded in a topological transformer. Instead of using a whole raw image as an input, extracted topological features can be sufficient to detect the treatment functionality. More complicated neural networks than vanilla CNN is also applicable.

Indeed, we can merge our two different approaches into a single, exclusive deep learning framework: detect interested cell organelles from a brightfield image and compute their topological properties. Even though our methodologies call for further pruning, but once completed, this elegant synthesis of state-of-the-art machinery and pure mathematics will be sensational in not only Alzheimer research but also general biomedical research and even further.


\newpage
\bibliography{main}        
\bibliographystyle{unsrt}  




\end{document}